\documentclass[aps,prb,twocolumn,english,showpacs,superscriptaddress,amssymb,amsfonts]{revtex4}
\usepackage[T1]{fontenc}
\usepackage[latin9]{inputenc}
\usepackage{amsmath}
\usepackage{dsfont}
\usepackage{amstext}

\usepackage{tocvsec2}

\usepackage{amssymb}
\usepackage{amsbsy}
\usepackage{amsthm}
\usepackage{epsfig}
\usepackage{graphicx}
\usepackage{bbm}
\usepackage{hyperref}
\usepackage{color}
\usepackage{multirow}
\usepackage{pdfsync}
\def\beq{\begin{equation}}
\def\eeq{\end{equation}}

\newcommand{\cs}{{C_6}}
\newcommand{\cf}{{C_4}}

\newcommand{\Ref}[1]{Ref.~\onlinecite{#1}}

\newcommand{\bss}{{\boldsymbol{\sigma}}}
\newcommand{\bst}{{\boldsymbol{T}}}

\newcommand{\bse}{{\boldsymbol{e}}}

\newcommand{\ie}{{\emph{i.e.~}}}
\makeatletter

\newcommand{\Rmnum}[1]{\expandafter\@slowromancap\romannumeral #1@}
\makeatother
\newcommand{\imth}{\hspace{1pt}\mathrm{i}\hspace{1pt}}

\newcommand{\eg}{{\emph{e.g.~}}}

\newcommand{\mbz}{{\mathbb{Z}}}
\newcommand{\bea}{\begin{eqnarray}}
\newcommand{\eea}{\end{eqnarray}}
\newcommand{\bpm}{\begin{pmatrix}}
\newcommand{\epm}{\end{pmatrix}}
\newcommand{\bal}{\begin{aligned}}
\newcommand{\eal}{\end{aligned}}

\newcommand{\dket}[1]{|{#1}\rangle}

\newcommand{\expval}[1]{\langle{#1}\rangle}
\newcommand{\spgs}{|\Psi_{\text{parton}}\rangle}
\newcommand{\ztsl}{$Z_2$ spin liquid}

\makeatother

\usepackage{babel}
\begin{document}
\title{Symmetry protected gapless $Z_2$ spin liquids}

\author{Yuan-Ming Lu}
\affiliation{Department of Physics, The Ohio State University, Columbus, OH 43210, USA}

\date{\today}

\begin{abstract}
Despite rapid progress in understanding gapped topological states, much less is known about gapless topological phases of matter, especially in strongly correlated electrons. In this work we discuss a large class of robust gapless quantum spin liquids in frustrated magnets made of half-integer spins, which are described by gapless fermionic spinons coupled to dynamical $Z_2$ gauge fields. Requiring $U(1)$ spin conservation, time reversal and certain space group symmetries, we show that certain spinon symmetry fractionalization class necessarily leads to a gapless spectrum. These gapless excitations are stable against any perturbations, as long as the required symmetries are preserved. Applying these gapless criteria to spin one-half systems on square, triangular and kagome lattices, we show that all gapped symmetric $Z_2$ spin liquids in Abrikosov-fermion representation can also be realized in Schwinger-boson representation. This leads to 64 gapped $Z_2$ spin liquids on square lattice, and 8 gapped states on both kagome and triangular lattices. 
\end{abstract}


\maketitle


\tableofcontents

\section{Introduction}

Tremendous progress has been made in understanding the interplay of symmetry and topology in gapped quantum many-body phases\cite{Senthil2015,Chiu2016,Wen2017} since the discovery of topological insulators\cite{Hasan2010,Hasan2011,Qi2011}. Take two spatial dimensions for example, a complete classification has been achieved\cite{Chen2013,Lu2012a,Bi2015,Essin2013,Mesaros2013,Hung2013,Lu2016a,Barkeshli2014,Teo2015a,Tarantino2016} for a generic gapped phase of interacting bosons that preserves certain (global) symmetries, characterized by possible fractional statistics and quantum numbers carried by elementary bulk and edge excitations therein. On the other hand, much less is known about gapless quantum phases beyond the free fermions regime\cite{Turner2013,Armitage2017}.

In this work we discuss a class of gapless and highly entangled quantum paramagnets in half-integer-spin systems, namely gapless \ztsl s\cite{Wen2002,Wen2002a}, whose low-energy physics is captured by an emergent $Z_2$ gauge field\cite{Read1991,Wen1991c} coupled to a pair superfluid (or superconductor) of fermionic ``spinons''. Unlike integer-spin magnon excitations, spinons carry a half-integer spin each, and form Kramers doublets as a projective representation of the time reversal symmetry. Therefore spinons are ``fractionalized'' excitations, each of which can be viewed as half of a magnon. Pairings between spinons can usually destroy the spinon fermi surface, leading to a \ztsl~with a gapped spectrum\cite{Wen2002}. Here we show that with time reversal and $U(1)$ spin rotational symmetries, certain space group symmetry implementations on spinons can give rise to a nodal superconductor of spinons with a gapless spectrum. These gapless \ztsl s are stable against any perturbations including arbitrary interactions between spinons, as long as the associated symmetries are preserved.

Between these spinon superconductors (\ztsl s) and usual electronic superconductors, there is a crucial difference on how space group symmetries can be implemented on spinons and on electrons. Since spinons are ``fractionalized'' collective excitations of the spin system, they can carry a fractional quantum number of spatial (and global) symmetries, a phenomena known as ``symmetry fractionalization''\cite{Essin2013}. The possible symmetry fractionalization classes of bosonic (or fermionic) spinons in a \ztsl~are classified by 2nd group cohomology $\mathcal{H}^2(G_s,Z_2)$ of the symmetry group\cite{Essin2013,Barkeshli2014} $G_s$. In this work, we show that certain fractionalization classes necessarily lead to gapless \ztsl s on a given lattice.

Previous studies on nodal superconductors\cite{Matsuura2013,Chiu2014} and gapless \ztsl s\cite{Wen2002a} are based on perturbative arguments at the free fermion level, \ie by analyzing symmetry-allowed mass terms for certain Dirac-like spectra. A natural step further is to address this question in a generic interacting system: can interactions between fermions (spinons) destroy a nodal superconductor (gapless \ztsl) in a non-perturbative fashion? In this work, by utilizing $U(1)$ spin rotational symmetry, we are able to establish the stability of gapless spinon spectrum non-perturbatively, based on properties of Schmidt decomposition\cite{Pollmann2010,Chen2011a,Watanabe2015} of a symmetric short-range-entangled ground state.

After establishing three symmetry criteria for gapless \ztsl s, we apply them to identify gapless \ztsl~s on square, triangular and kagome lattices. They also enable us to identify the gapped symmetric \ztsl s on these lattices, where the kagome lattice case is experimentally relevant to the possibly gapped paramagnetic ground state of Herbertsmithite\cite{Mendels2007,Fu2015}. We found that on all three lattices, the only possible gapped symmetric \ztsl s are those realized in Schwinger-boson representation\cite{Sachdev1992,Wang2006,Yang2016}.

This work is organized as follows. In section \ref{PARTON CONSTRUCTION} we setup a generalized parton construction (\ref{general parton construction}) for a half-integer spin system, using which the \ztsl s will be constructed and investigated in this paper. This representation reduces to the Schwinger-boson\cite{Schwinger1965} and Abrikosov-fermion representation\cite{Abrikosov1965} in the spin-$\frac12$ case. In section \ref{GAPLESS CRITERIA} we establish three symmetry criteria (\ref{criterion 1: two mirrors}), (\ref{criterion 2:mirror+TRS}) and (\ref{criterion 3:n=odd})-(\ref{criterion 3:n=even}) for gapless \ztsl s, by supplementing intuitive single-particle proofs with non-perturbative arguments with respect to interactions. Next we apply these criteria to symmetric \ztsl s on square (section \ref{SQUARE}), triangular (section \ref{TRIANGLE}) and kagome (section \ref{KAGOME}) lattices, to identify the gapped and gapless \ztsl s on these lattices. Finally we conclude in section \ref{SUMMARY}.

\section{Parton construction for symmetric $Z_2$ spin liquids of half-integer spins}\label{PARTON CONSTRUCTION}

\subsection{A general parton construction for half-integer spins}

In this section we setup a general parton construction for half-integer spins, which allows us to construct an arbitrary ground state in a half-integer-spin system. In particular in the case of spin-$\frac12$ system, it becomes the familiar Schwinger-boson\cite{Schwinger1965} or Abrikosov-fermion\cite{Abrikosov1965} representation. In fact this parton construction also applies to integer spins, such as spin-1 systems\cite{Xu2012,Wang2011c,Zheng2015}. However in this work, we will focus on half-integer spins which give rise to gapless \ztsl s. Hereafter we will always assume $2S=1\mod2$ unless specified otherwise.

In this construction, each spin-$S$ ($2S\in\mbz$) operator $\vec{\bf S}_{\bf r}$ is represented by $2S$ pairs of spin-$\frac12$ fermionic (or bosonic) ``partons'' $\{\phi_{{\bf r},a,\alpha}|1\leq a\leq2S,\alpha=\uparrow,\downarrow\}$, where ${\bf r}$ label a specific lattice site:
\bea\label{general parton construction}
\vec{\bf S}_{\bf r}=\frac12\sum_{a=1}^{2S}\sum_{\alpha,\beta=\uparrow,\downarrow}\phi^\dagger_{{\bf r},a,\alpha}\vec{\boldsymbol\sigma}_{\alpha,\beta}\phi_{{\bf r},a,\beta},~~~2S\in\mbz.
\eea
and $\vec{\boldsymbol\sigma}$ represent the three Pauli matrices. $a$ is the flavor index and $\alpha,\beta$ are spin indices. These partons physically correspond to spinon excitations in a quantum spin liquid, each of which carries spin-$1/2$ and can be viewed as half of a magnon.

Following the rule for addition of angular momentum, to ensure that $\vec{\bf S}_{\bf r}$ represents a half-integer spin we require that $2S$ to be an odd integer and
\bea\label{cond:odd spinons per site}
\sum_{a,\alpha}\phi^\dagger_{{\bf r},a,\alpha}\phi_{{\bf r},a,\alpha}=2S,~~~\forall~{\bf r}.
\eea
Note that we have not specify the statistics of spinons $\{\phi_{{\bf r},a,\alpha}\}$. Bose versus Fermi statistics of partons will lead to two different parton constructions for the same spin-$S$ system, such as Schwinger-boson versus Abrikosov-fermion representation for the $S=\frac12$ case.

Once the parton Hamiltonian $\hat H_\text{parton}$ for $\{\phi_{{\bf r},a,\alpha}\}$ is constructed and its ground state $\dket{\Phi_\text{parton}}$ obtained, the many-spin state $\dket{\Psi_\text{spin}}$ is achieved by implementing certain projection operators ($\hat P_{\bf r}$ on site ${\bf r}$) on the parton ground state
\bea\label{spin wf and projector}
\dket{\Psi_\text{spin}}=(\prod_{\bf r}\hat P_{\bf r})\dket{\Phi_\text{parton}}
\eea
This is because the Hilbert space of partons $\{\phi_{{\bf r},a,\alpha}|1\leq a\leq2S,\alpha=\uparrow,\downarrow\}$ per site is larger than the $(2S+1)$-dimensional spin-$S$ Hilbert space, and hence we need to impose certain constraints on the parton Hilbert space to reduce it to the physical spin-$S$ Hilbert space. For example, condition (\ref{cond:odd spinons per site}) is one necessary constraint for a general parton construction (\ref{general parton construction}) of spin-$S$ systems. All the constraints are enforced by a projection operator $\{\hat P_{\bf r}\}$ in (\ref{spin wf and projector}) to produce a physical spin-$S$ wavefunction.

In the $S=1/2$ case, the generalized parton construction (\ref{general parton construction}) reduces to the widely-adopted Abrikosov-fermion\cite{Abrikosov1965,Affleck1988b} or Schwinger-boson\cite{Schwinger1965} representation, depending on whether partons $\{\phi_{{\bf r},a,\alpha}\}$ obey Fermi or Bose statistics. The on-site constraint for $S=1/2$ case becomes the single-occupancy condition
\bea
S=\frac12\Longrightarrow\sum_{\alpha}\phi^\dagger_{{\bf r},\alpha}\phi_{{\bf r},\alpha}=1,~~~\forall~{\bf r}.
\eea
and the associated projection operator is nothing but the well-known Gutzwiller projector
\bea
\prod_{\bf r}\hat P_{\bf r}(S=\frac12)=\prod_{\bf r}\big(1-\sum_{\alpha}\phi^\dagger_{{\bf r},\alpha}\phi_{{\bf r},\alpha}\big)\equiv\hat P_G
\eea
In the $2S>1$ case, the onsite constraints are typically more complicated\cite{Xu2012,Wang2011c,Zheng2015} than condition (\ref{cond:odd spinons per site}). However for the purpose of this work, it turns out that constraint (\ref{cond:odd spinons per site}) itself suffices to establish the stability for gapless \ztsl s of half-integer spins. Therefore we will not bother to explicitly write down the general constraints for the parton construction (\ref{general parton construction}) of an arbitrary spin-$S$ system.


\subsection{A brief review on symmetry fractionalization in $Z_2$ spin liquids}

Now that the physical many-spin wavefunction $\dket{\Psi_\text{spin}}$ is related to the many-body state $\dket{\Phi_\text{parton}}$ of partons $\{\phi_{{\bf r},a,\alpha}\}$ by the projection (\ref{spin wf and projector}), how different are these two wavefunctions? Do they share the same symmetry properties? Take \ztsl s for example, in order to obtain a spin wavefunction $\dket{\Psi_\text{spin}}$ for a \ztsl~that preserves a symmetry group $G_s$, what are the symmetries required for the corresponding parton Hamiltonian $\hat H_\text{parton}$ and associated parton state $\dket{\Phi_\text{parton}}$?

In the general parton construction (\ref{general parton construction}), it is easy to notice that both the physical spin operators (\ref{general parton construction}) and the onsite constraints (\ref{cond:odd spinons per site}) are invariant under certain gauge transformations on partons $\{\phi_{{\bf r},a,\alpha}\}$. Two well-known examples are the $SU(2)$ gauge transformations\cite{Affleck1988b} $\{W_{\bf r}\in SU(2)\}$ for Abrikosov-fermion representation of spin-$1/2$:
\bea\notag
&\bpm\phi_{{\bf r},\uparrow}\\ \phi^\dagger_{{\bf r},\downarrow}\epm\longrightarrow W_{\bf r}
\bpm\phi_{{\bf r},\uparrow}\\ \phi^\dagger_{{\bf r},\downarrow}\epm,~~~W_{\bf r}=e^{\imth\theta_{\bf r}\hat{\bf n}_{\bf r}\cdot\vec\sigma}\in SU(2),\\
&S=1/2,~~~\{\phi_{{\bf r},\alpha}\}~\text{are Abrikosov fermions}.
\eea
and the $U(1)$ gauge transformations\cite{Sachdev1992,Wang2006} $\{e^{\imth\theta_{\bf r}}\in U(1)\}$ for Schwinger-boson representation of spin-$1/2$:
\bea
&\notag\phi_{{\bf r},\alpha}\longrightarrow e^{\imth\theta_{\bf r}}\phi_{{\bf r},\alpha},~~~e^{\imth\theta_{\bf r}}\in U(1),\\
&S=1/2,~~~\{\phi_{{\bf r},\alpha}\}~\text{are Schwinger bosons}.
\eea
For a given parton construction, these gauge transformations are featured by a gauge group $G_g^{(0)}$, which \eg is $SU(2)$ for spin-$1/2$ Abrikosov-fermion case and $U(1)$ for spin-$1/2$ Schwinger-boson case.

As a consequence of these gauge redundancies, the parton state $\dket{\Phi_\text{parton}}$ may not faithfully preserve the symmetry group $G_s$ of the physical spin wavefunction $\dket{\Psi_\text{spin}}$. Most generally, the symmetry implementations on partons form a projective representation\cite{Wen2002,Essin2013} of the symmetry group $G_s$. In the case of a \ztsl, the original gauge group $G_g^{(0)}$ is broken down to a subgroup (labeled ``invariant gauge group'' or IGG\cite{Wen2002}) IGG~$=Z_2$. And the different projective representations of the symmetry group $G_s$ is mathematically classified by the 2nd group cohomology $\mathcal{H}^2(G_s,\text{IGG}=Z_2)$.

More concretely, we demonstrate the projective implementation of a physical symmetry $g\in G_s$ in the general parton construction (\ref{general parton construction}). The symmetry action $U_g$ on partons $\{\phi_{{\bf r},a,\alpha}\}$ is generally a combination of physical symmetry $g$ and its associated gauge rotation $\{W_g({\bf r})\in G_g\}$:
\bea
&\hat U_g\equiv\prod_{\bf r}W_g({\bf r})\cdot g,\\
\notag&\hat U_g\phi_{{\bf r},a,\alpha}\hat U_g^{-1}=W_g(\hat g{\bf r})\phi_{\hat g{\bf r},a,\hat g\alpha}W_g^{-1}(\hat g{\bf r}).
\eea
While physical symmetries $g,h$ and their product $gh\equiv g\cdot h$, their associated implementations on partons satisfy the following condition
\bea
\hat U_g\cdot\hat U_h=\hat\Omega(g,h)\cdot\hat U_{gh},~~~\hat\Omega(g,h)\in\text{IGG}.
\eea
The symmetry implementations $\{\hat U_g|g\in G_s\}$ on partons $\{\phi_{{\bf r},a,\alpha}\}$ form a so-called ``projective symmetry group'' or PSG\cite{Wen2002}, which is an extension of symmetry group $G_s$ satisfying\cite{Wen2002}
\bea
\text{PSG}/G_s=\text{IGG}.
\eea
In the case of \ztsl s, the Invariant Gauge Group takes a particularly simple form ($\hat N_\phi$ is the total number of partons/spinons)
\bea
\text{IGG}=\{(-1)^{\hat N_\phi},\hat 1\}\simeq Z_2,~~~\hat N_\phi\equiv\sum_{{\bf r},a,\alpha}\phi^\dagger_{{\bf r},a,\alpha}\phi_{{\bf r},a,\alpha}.\label{parton number}
\eea
and therefore
\bea
&\notag\hat\Omega(g,h)\phi_{{\bf r},a,\alpha}\hat\Omega^{-1}(g,h)=\omega(g,h)\phi_{{\bf r},a,\alpha},\\
&\{\omega(g,h)=\pm1|g,h\in G_s\}\in\mathcal{H}^2(G_s,Z_2).\label{psg:phase factors}
\eea
Clearly operations within IGG commute with all symmetry implementations $\{\hat U_g|g\in G_s\}$ on partons
\bea
[\hat\Omega(g,h),\hat U_{g^\prime}]=0,~~~\forall~g,h,g^\prime\in G_s.
\eea
In this case, the projective symmetry implementations $\{\hat U_g|g\in G_s\}$ on partons $\{\phi_{{\bf r},a,\alpha}\}$ form a central extension of symmetry group $G_s$. Different symmetric \ztsl s preserving the same symmetry group $G_s$ are hence classified by different $\pm1$-valued phase factor sets $\{\omega(g,h)=\pm1|g,h\in G_s\}$, which mathematically corresponds to the classification of 2nd group cohomology $\mathcal{H}^2(G_s,Z_2)$ as in (\ref{psg:phase factors}). These phase factors are also subject to associativity
\bea
\omega(g,h)\omega(f,gh)=\omega(fg,h)\omega(f,g)
\eea
They are well-defined only up gauge transformations
\bea
\omega(g,h)\rightarrow\omega(g,h)\frac{w_{gh}}{w_g\cdot w_h},~~~w_g,w_h,w_{gh}=\pm1.
\eea
Two sets of phase factors related by a gauge transformation belong to the same fractionalization class and describe the same \ztsl. Therefore gauge-inequivalent phase factor sets $\{\omega(g,h)=\pm1|g,h\in G_s\}$ correspond to different fractionalization classes, and hence distinct symmetric \ztsl s.

If all phase factors can be chosen to be trivial \ie $\omega(g,h)\equiv1$, then the physical symmetry group $G_s$ is implemented faithfully on partons. On other other hand, if it is impossible to trivialize all phase factors by any gauge transformation, there exists at least two elements $g,h\in G_s$ such that $U_g\cdot U_h=(-1)^{\hat N_\phi}\cdot U_{gh}$. These
symmetries $g,h$ act projectively on spinons/partons $\{\phi_{{\bf r},a,\alpha}\}$, and this phenomenon is coined ``symmetry fractionalization''\cite{Essin2013,Barkeshli2014,Tarantino2016} in quantum spin liquids. For gapped symmetric topological orders\cite{Wen2004B}, different symmetry fractionalization classes of anyon excitations therein can be fully classified in an abstract setup under the framework of unitary modular tensor category\cite{Barkeshli2014,Tarantino2016}. The parton construction discussed here not only provides a physical manifestation for the abstract mathematical classification, but also applies to gapless \ztsl s\cite{Wen2002,Wen2002a} which are beyond the current categorical framework.

Below we list a few gauge-invariant phases, whose different values will correspond to distinct symmetric \ztsl s. Each gauge-invariant phase is typically associated with a series of symmetry operations which yields the identity operation $\bse$ in the symmetry group $G_s$, as shown in the left column of TABLE \ref{tab:square}-\ref{tab:kagome}.

For time reversal symmetry $\bst$ we have
\bea
\bst^2=(-1)^{\hat N_\phi}\Longrightarrow\omega_\bst\equiv\omega(\bst,\bst)=-1
\eea
since partons/spinons are all Kramers doublets. $\bse$ is the identity element of symmetry group $G_s$.

For mirror reflection symmetry $\bss$ we have
\bea
\bss^2=\bse\Longrightarrow\omega_\bss\equiv\omega(\bss,\bss)=\pm1.
\eea

For $n$-fold rotational symmetry $C_n$ we have
\bea
(C_n)^n=\bse\Longrightarrow\omega_{C_n}\equiv\prod_{i=1}^{n-1}\omega\big(C_n,(C_n)^i\big)=\pm1.
\eea

With both time reversal and mirror reflection, we have
\bea
\bss^{-1}\bst^{-1}\bss\bst=\bse\Longrightarrow\omega_{\bss\bst}\equiv\frac{\omega(\bss,\bst)}{\omega(\bst,\bss)}=\pm1.
\eea


These gauge invariant phase factors will be used in following discussions on the gapless criteria and applications to \ztsl s on various lattices.

\subsection{Spinon pair superfluid description of $Z_2$ spin liquids}

In the following, we setup the most generic parton Hamiltonian $\hat H_\text{parton}$ for a symmetric \ztsl, in the framework of parton construction (\ref{general parton construction}). The symmetry group $G_s$ considered here includes at least $U(1)$ spin rotational symmetry (say, along $\hat z$-axis), time-reversal symmetry $\bst$, and certain space group symmetries:
\bea\label{sym group}
G_s\supset U(1)_{{\bf S}^z}\times Z_2^\bst\times\text{Space group}
\eea
We do not consider spin-orbit couplings in the system, so the space group symmetries considered here are pure spatial operations involving no spin rotations.

Under time reversal operation $\hat\bst=U_\bst\cdot\mathcal{K}$ where $\mathcal{K}$ represents complex conjugation, the partons/spinons transform as spin-$1/2$ Kramers doublets by construction in (\ref{general parton construction}):
\bea\label{sym:time reversal}
\hat\bst \phi_{{\bf r},a,\alpha} \hat\bst^{-1}=\sum_\beta\imth(\sigma_y)_{\alpha,\beta}\phi_{{\bf r},a,\beta}
\eea
Under $U(1)_{{\bf S}^z}$ spin rotation they transform as
\bea\label{sym:spin rotation S^z}
e^{\imth\theta\hat{\bf S}^z}\phi_{{\bf r},a,\alpha}e^{-\imth\theta\hat{\bf S}^z}=\sum_\beta e^{\imth\frac{\theta}2(\sigma_z)_{\alpha,\beta}}\phi_{{\bf r},a,\beta}
\eea

The conservation of $\hat z$-component of total spins makes the Nambus basis $\psi_{\bf r}\equiv(\phi_{{\bf r},a,\uparrow}\,\phi^\dagger_{{\bf r},a,\downarrow})^T$ a convenient choice to diagonalize the parton Hamiltonian. Most generally, the symmetry-allowed parton Hamiltonian of a spin-$S$ system can be written in the Nambu basis as
\bea
\notag&\hat H_\text{parton}=\hat H_{MF}+\hat H_{int},\\
&\hat H_{MF}=\sum_{\bf r,r^\prime}\psi^\dagger_{\bf r}\expval{{\bf r}|{\bf r}^\prime}\psi_{\bf r^\prime},~~~\psi_{\bf r}\equiv\bpm\phi_{{\bf r},a,\uparrow}\\ \phi^\dagger_{{\bf r},a,\downarrow}\epm.\label{nambu basis}
\eea
where $\hat H_{MF}$ represents the quadratic part at mean-field level and $\hat H_{int}$ for generic symmetry-preserving interactions between partons. The $(4S)\times(4S)$ matrices $\expval{\bf r|r^\prime}=\expval{\bf r^\prime|r}^\dagger$ are mean-field amplitudes for spinons. The spin rotational symmetry (\ref{sym:spin rotation S^z}) is nothing but the global phase rotation in the Nambu basis:
\bea
e^{\imth\theta\hat{\bf S}^z}\psi_{\bf r}e^{-\imth\theta\hat{\bf S}^z}=e^{\imth\frac\theta2}\psi_{\bf r},
\eea
while time reversal symmetry (TRS) plays the role of an \emph{anti-unitary particle-hole symmetry} on the Nambu spinor:
\bea
\notag&\hat\bst\psi_{\bf r}\hat\bst^{-1}=\bpm\phi_{{\bf r},a,\downarrow}\\-\phi^\dagger_{{\bf r},a,\uparrow}\epm=\imth\tau_y\psi_{\bf r}^\ast,\\
&\hat\bst^2=(-1)^{\hat N_\phi}=(-1)^{2NS+\hat F},~~~\hat F\equiv\sum_{\bf r}\psi^\dagger_{\bf r}\psi_{\bf r}.\label{nambu:time reversal}
\eea
where $\vec\tau$ are Pauli matrices for the Nambu index, and $N$ is the total number of lattice sites.

For a symmetric \ztsl, the time-reversal-symmetric pair superfluid Hamiltonian in the partons/spinon $\{\phi_{\bf r}\}$ basis is mapped to a Bloch Hamiltonian of particle-hole-symmetric insulator in the Nambu basis $\{\psi_{\bf r}\}$. Due to $U(1)_{{\bf S}^z}$ spin conservation, the Nambu particles $\{\psi_{\bf r}\}$ have a conserved particle number $\hat F$ as defined in (\ref{nambu:time reversal}). Although usually $(-1)^{\hat F}$ is used to denote fermion parity, here we use it for both (hard-core) Bose and Fermi statistics of partons $\{\phi_{{\bf r},a,\alpha}\}$. It is instructive to compare the vacua and ground states in these two different basis. Denoting the Fock vacuum of partons $\{\phi_{\bf r}\}$ as $\dket{0}$ with
\bea\label{Fock vacuum}
\phi_{{\bf r},a,\alpha}\dket{0}=0,~~~\forall~{\bf r},a,\alpha.
\eea
the vacuum $\dket{empty}$ in the Nambu basis is given by
\bea
&\notag\psi_{\bf r}\dket{empty}=0\Longrightarrow\\
&\dket{empty}=(\prod_{{\bf r},a}\phi^\dagger_{{\bf r},a,\downarrow})\dket{0}=\dket{-S,-S,\cdots}\label{nambu vacuum}
\eea
This is nothing but the fully polarized state where all spins point to negative $\hat z$-direction. Now that partons $\{\psi_{\bf r}\}$ (or $\{\phi_{{\bf r}}\}$) are either hard-core bosons or fermions, there is also a fully filled state in the Nambu basis
\bea
\notag&\psi_{\bf r}^\dagger\dket{full}=0\Longrightarrow\dket{full}=(\prod_{{\bf r},a}\psi^\dagger_{{\bf r},a,1}\psi^\dagger_{{\bf r},a,2})\dket{empty}\\
&=\prod_{{\bf r},a}\phi_{{\bf r},a,\downarrow}^\dagger\dket{0}=\dket{+S,+S,\cdots}.\label{nambu full}
\eea
It is the other polarized state where all spins point to the positive $\hat z$-direction, and the time reversal partner of the Nambu vacuum
\bea
\dket{full}=\hat\bst\dket{empty}.
\eea
It's straightforward to show that no matter partons $\{\psi_{\bf r}\}$ are either hard-core bosons or fermions, the Nambu particle number $\hat F$ in (\ref{nambu:time reversal}) is related to the total spin $\hat{\bf S}^z$ by
\bea
\hat2{\bf S}^z\equiv\sum_{{\bf r},a}\big(\phi^\dagger_{{\bf r},a,\uparrow}\phi_{{\bf r},a,\uparrow}-\phi^\dagger_{{\bf r},a,\downarrow}\phi_{{\bf r},a,\downarrow}\big)=\hat F-2NS.
\eea
Consequently, TRS (\ref{nambu:time reversal}) enforces the Nambu particle number $\hat F$ to be at ``half filling'':
\bea\label{nambu:filling}
{\bf S}^z=0\Longrightarrow\frac{\hat F}N=\frac1N\sum_{\bf r}\psi_{\bf r}^\dagger\psi_{\bf r}=2S=\text{odd}.
\eea
Hereafter we'll stick to the Nambu basis which allows us to make the best of $U(1)_{{\bf S}^z}$ symmetry.\\

A generic time-reversal-symmetric \ztsl~can be obtained by projecting the following parton state
\begin{align}
\notag\dket{\Psi_\text{parton}}=\sum_{\{{\bf r}_i,a_i,b_i\}}\Psi(\{{\bf r}_i,a_i,b_i\})\prod_{i=1}^{NS}\psi^\dagger_{{\bf r}_i,a_i,1}\psi^\dagger_{{\bf r}_i,b_i,2}\dket{empty}.&\\
 \label{parton wf:z2 sl}&
\end{align}
which satisfies
\bea
{\bf S}^z\dket{\Psi_\text{parton}}=0.
\eea
apart from one exception described below. In a system with $N=$~odd half-integer spins, Kramers theorem dictates at least two-fold degeneracy for all eigenstates. Indeed the parton state (\ref{parton wf:z2 sl}) will not be well-defined since $NS$ is a half integer. In this case, the parton ground states typically create $NS\pm\frac12$ pairs of Nambu particles on top of Nambu vacuum $\dket{empty}$, with a total spin ${\bf S}^z=\pm\frac12$.

Next we analyze the symmetry implementations on the Nambu vacuum $\dket{empty}$ and parton state $\dket{\Psi_\text{parton}}$. Since the Nambu vacuum $\dket{empty}$ in (\ref{nambu vacuum}) is a physical state where all spins are fully polarized along negative $\hat z$ direction, symmetries must act faithfully on the Nambu vacuum\footnote{In comparison, the symmetry implementations on the parton Fock vacuum $\dket{0}$ defined in (\ref{Fock vacuum}) are quite different: $\hat U_g\cdot\hat U_h\dket{0}=\big[\omega(g,h)\big]^{2NS}\hat U_{gh}\dket{0}$ for all $g,h\in G_s$, with the only exception that $\hat\bst^2\dket{0}=\dket{0}$.} \ie
\bea
\hat U_g\cdot\hat U_h\dket{empty}=\hat U_{gh}\dket{empty},~\forall~g,h\in G_s.
\eea
with the only exception that
\bea
\hat\bst^2\dket{empty}=(-1)^{2NS}\dket{empty}.
\eea
associated with Kramers degeneracy if $2NS=$~odd. In other words, the series of combined symmetry operations on the left column of TABLE \ref{tab:square}-\ref{tab:kagome} all act trivially on the Nambu vacuum (\ref{nambu vacuum}), except for $\bst^2=\bse$.

On an arbitrary parton state $\dket{\Psi_\text{parton}}$ in (\ref{parton wf:z2 sl}), symmetry operations act in the following way
\bea
\hat U_g\cdot\hat U_h\dket{\Psi_\text{parton}}=\big[\omega(g,h)\big]^{\hat F}\cdot\hat U_{gh}\dket{\Psi_\text{parton}}
\eea
or equivalently
\bea\label{proj symm:nambu spinon}
\hat U_g\cdot\hat U_h=\big[\omega(g,h)\big]^{\hat F}\cdot\hat U_{gh}.
\eea
This projective symmetry operation is crucial for the gapless criteria for \ztsl s as we will show below.

\section{Criteria for symmetry protected gapless $Z_2$ spin liquids}\label{GAPLESS CRITERIA}

Previously we have setup the parton Hamiltonian of a generic symmetric pair superfluid,  which describes a symmetric $Z_2$ spin liquid in the parton construction (\ref{general parton construction}), irrespective of the (hard-core)Bose or Fermi statistics of partons. An important property of a \ztsl~is whether there is a finite energy gap in the excitation spectrum or not. Usually the existence of an excitation gap depends on the specified Hamiltonian of the system. However in certain quantum spin liquids, one can sometimes rule out the possibility of an energy gap once the physical Hilbert space and symmetry implementations on spinons are specified\cite{Oshikawa2006,Zaletel2015,Cheng2016}, without referring to any specific Hamiltonian.

Here we will address the following question in the context of symmetric \ztsl s: given a lattice system of half-integer spins with certain space group symmetries, for a specific symmetry fractionalization class of spinons, is it possible to construct a gapped symmetric \ztsl? Always assuming time reversal symmetry (\ref{sym:time reversal}) and $U(1)$ spin rotational symmetry (\ref{sym:spin rotation S^z}), we consider all possible fractionalization classes of crystal symmetry on spinons in a \ztsl~(see \eg TABLE \ref{tab:square}-\ref{tab:kagome}). We do not specify the statistics of spinons $\{\phi_{{\bf r},a,\alpha}\}$, \ie they can be either hard-core bosons or fermions. If the generic parton/spinon Hamiltonian (\ref{nambu basis}) cannot support a unique gapped ground state, our parton construction will lead to a gapless \ztsl, whose low-energy physics are described by gapless spinons coupled to dynamical $Z_2$ gauge fields. Previously, gapless \ztsl s were mostly discussed at the mean-field level\cite{Wen2002,Wen2002a} in the parton construction, but it is not clear whether interactions between spinons or gauge fluctuations can open up a gap in the spectrum. Although gapless \ztsl s were found numerically in projected wavefunctions\cite{Capriotti2001,Hu2013a} on the square lattice, the stability of these states with respect to general perturbations remains an open issue. Here using the conserved particle number $\hat F$ in the Nambu basis, as a consequence of $U(1)_{{\bf S}^z}$ symmetry, we are able to non-perturbatively prove the stability of certain gapless \ztsl s with respect to any symmetry-preserving perturbations, including arbitrary interactions between spinons.

In the following we establish three criteria on the symmetry fractionalization class of spinons, which lead to a gapless spectrum of the interacting parton/spinon Hamiltonian (\ref{nambu basis}). These criteria dictate the stability of gapless \ztsl s associated with certain spinon fractionalization class $\{\omega(g,h)|g,h\in G_s\}\in\mathcal{H}^2(G_s,Z_2)$. Independent of whether partons/spinons are hard-core bosons or fermions, we take the following strategy to establish the gapless criteria. Given certain spinon symmetry fractionalization class $\{\omega(g,h)|g,h\in G_s\}$, we first prove symmetry-enforced zero-energy degeneracy, at the level of bilinear mean-field Ansatz $\hat H_{MF}$ of partons/spinons. Next we go beyond mean-field Ansatz, and argue the impossibility of a unique gapped many-body ground state for the interacting parton/spinon Hamiltonian (\ref{nambu basis}), using the Schmidt decomposition of a short-range-entangled (gapped) state\cite{Pollmann2010,Chen2011a,Watanabe2015}.

\begin{table}[tb!]
\begin{tabular} {|c|c|c|c|}
\hline
Algebraic Identity&\multirow{2}{1.3cm}{SB $b_{\alpha}$ in \Ref{Yang2016}} &
\multirow{2}{1.3cm}{AF $f_{\alpha}$ in \Ref{Wen2002}}&\multirow{2}{2.3cm}{fractionalization class of spinons}\\
&&&\\ \hline
$T^{-1}_{2}T^{-1}_{1}T_{2}T_{1}=\bse$&(-1)$^{p_1}$&$\eta_{xy}$&$\omega_{T_1T_2}$\\ \hline
$\bss^{-1}T_1\bss T_1^{-1}=\bse$&(-1)$^{p_2}$&$\eta_{xpy}$&$\omega_{\bss T_1}$\\ \hline
$(\bss T_2)^2=\bse$&(-1)$^{p_3+p_4}$&$\eta_\bss\eta_{xpx}$&$\omega_{\bss T_2}\omega_\bss$\\ \hline
$\cf^{-1}T_1\cf T_2=\bse$&1&1&1\\ \hline
$\cf^{-1}T_2\cf T_1^{-1}=\bse$&(-1)$^{p_2+p_3}$&$\eta_{xpy}\eta_{xpx}$&$\omega_{\bss T_1}\omega_{\bss T_2}$\\ \hline
$\bss^2=\bse$&(-1)$^{p_4}$&$\eta_\bss$&$\omega_\bss$\\ \hline
$R_{xy}^{2}=(\cf\bss)^2=\bse$&(-1)$^{p_4+p_7}$&$\eta_\bss\eta_{\bss\cf}$&$\omega_{R}$\\ \hline
$(\cf)^{4}=\bse$&1&$\eta_\cf$&$\omega_{\cf}$\\ \hline
$T_1^{-1}\bst^{-1}T_1\bst=\bse$&(-1)$^{p_8}$&$\eta_{t}$&$\omega_{T_1\bst}$\\ \hline
$T_2^{-1}\bst^{-1}T_2\bst=\bse$&(-1)$^{p_8}$&$\eta_{t}$&$\omega_{T_1\bst}$\\ \hline
$\bss^{-1}\bst^{-1}\bss\bst=\bse$&(-1)$^{p_4}$&$\eta_{\bss\bst}$&$\omega_{\bss\bst}$\\ \hline
$R_{xy}^{-1}\bst^{-1}R_{xy}\bst=\bse$&(-1)$^{p_4+p_7}$&$\eta_{\cf\bst}\eta_{\bss\bst}$&$\omega_{R\bst}$\\ \hline
$\bst^2=\bse$&-1&-1&-1\\ \hline
\end{tabular}
\caption{Square lattice: spinon symmetry fractionalization class $(\mbz_2)^9\subset\mathcal{H}^2(P4gm\times Z_2^\bst,\mbz_2)$ for $Z_2$ spin liquids of half-integer spins, and their realizations in $S=\frac12$ Schwinger-boson (SB) and Abrikosov-fermion (AF) representations, following the notation of \Ref{Wen2002,Yang2016}. The fractionalization classes from 2nd group cohomology\cite{Essin2013} are characterized by nine $Z_2$-valued gauge-invariant phases $\omega=\pm1$. We show that a $Z_2$ spin liquid is gapless if violating any of the following conditions: $\omega_{\bss}=\omega_{\bss\bst}$, $\omega_{R}=\omega_{R\bst}$ and $\omega_{\cf}=1$. This leads to only $2^6=64$ distinct gapped $Z_2$ spin liquids, all realized in both SB\cite{Yang2016} and AF\cite{Wen2002} representations. There are also 96 symmetry protected gapless \ztsl s in the AF representation.}
\label{tab:square}
\end{table}

\begin{figure}
\includegraphics[width=0.9\columnwidth]{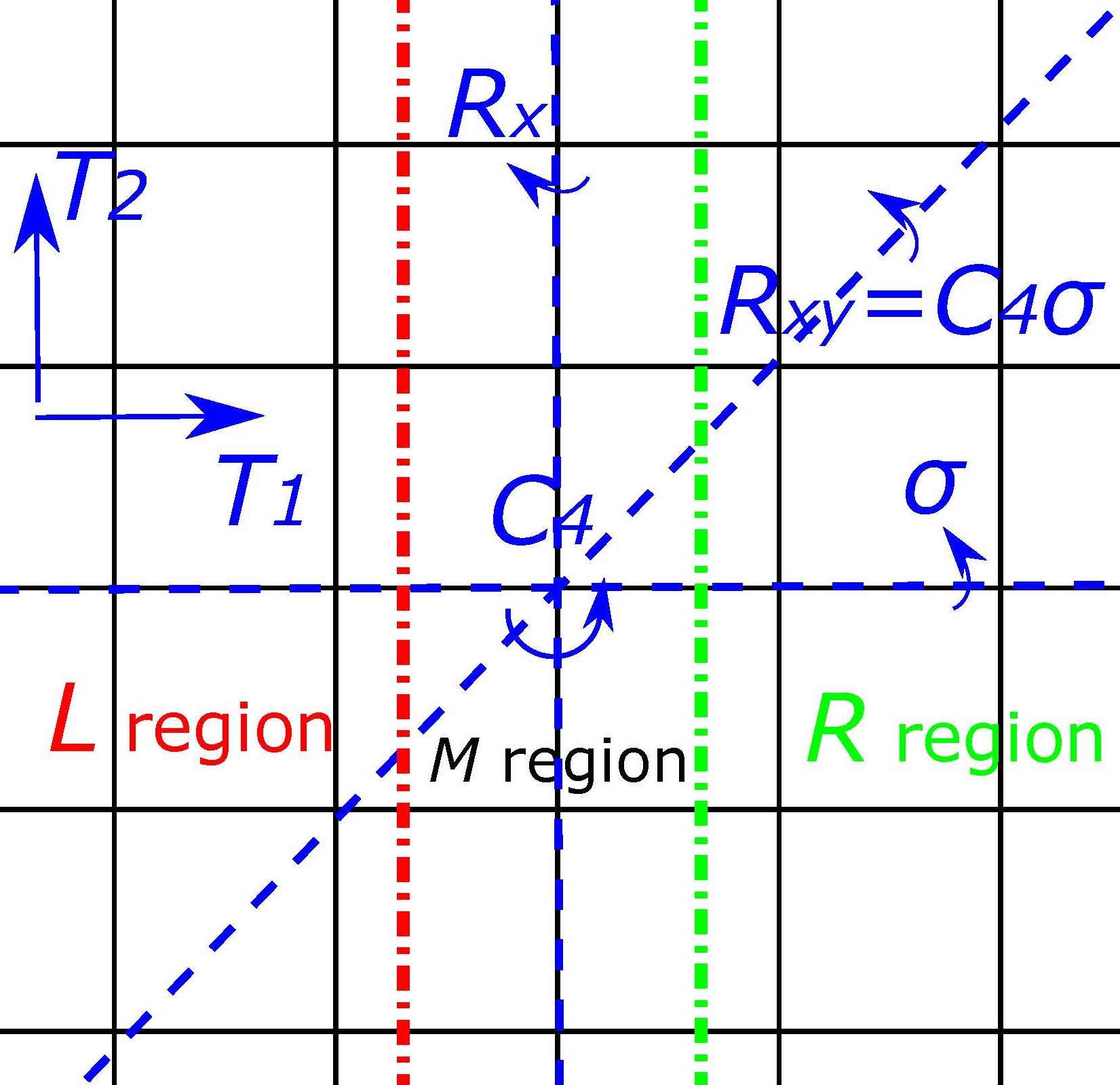}
\caption{(color online) Space group ($P4gm$) symmetries of square lattice, generated by two translations $T_{1,2}$, mirror reflection $\bss$ and 4-fold rotation $\cf$. Two other mirror reflections are defined as $R_{xy}=\cf\bss$ and $R_x=(\cf)^2\bss$. Red and green dotted dash lines denote two entanglement cuts of the lattice, related by mirror $R_x$ or inversion $I\equiv C_2=(\cf)^2$. }
\label{fig:square}
\end{figure}

\subsection{Two perpendicular mirror planes intersecting at one half-integer spin}

The first criterion comes from two perpendicular mirror reflection planes, which intersect at one lattice site with a half-integer spin on it. Take square lattice for example, we consider two perpendicular mirror planes $\bss$ and $R_x$ in FIG. \ref{fig:square}. Consider the following projective symmetry implementation on spinons
\bea\notag
&\hat\bss\hat R_x\hat\bss^{-1}\hat R_x^{-1}=\bse\Longrightarrow\\
\label{criterion 1: two mirrors}&\frac{\omega(\bss,R_x)}{\omega(R_x,\bss)}=-1\Leftrightarrow
\hat U_\bss\hat U_{R_x}\hat U_\bss^{-1}\hat U_{R_x}^{-1}=(-1)^{\hat F}.
\eea
where $\hat F=\sum_{\bf r}\psi^\dagger_{\bf r}\psi_{\bf r}$ is the parton number in the Nambu basis. Since the combination of the two mirror reflections is nothing but the two-fold rotation $C_{2}=\bss R_x$ around their intersection site, the above gauge-invariant phase factor can also be written as
\bea
\frac{\omega(\bss,R_x)}{\omega(R_x,\bss)}=\frac{\omega(\bss R_x,\bss R_x)}{\omega(R_x,R_x)\cdot\omega(\bss,\bss)}\equiv\frac{\omega_{C_2}}{\omega_{R_x}\cdot\omega_\bss}=-1.
\eea

First it's straightforward to show that in mean-field Ansatz, each single-particle level of partons/spinons must be at least 2-fold degenerate. The $4NS\times4NS$ mean-field Ansatz Hamiltonian can be diagonalized as
\bea\label{mean-field ham diagonalized}
\hat H_{MF}=\sum_{\bf r,r^\prime}\psi^\dagger_{\bf r}\expval{{\bf r}|{\bf r}^\prime}\psi_{\bf r^\prime}=\sum_E E\gamma_E^\dagger\gamma_E,
\eea
where $E$ represents the single-particle eigen-energy, and $\gamma_E$ is the annihilation operator of the associated eigenstate.
TRS in (\ref{nambu:time reversal}) as an anti-unitary particle-hole symmetry indicates that single-spinon levels always show up in pairs with opposite energies:
\bea
\hat\bst\gamma_E\hat\bst^{-1}=\gamma_{-E}^\dagger.
\eea
As a result, the number of positive and negative single-spinon energy levels must both equal $2NS$.

In the presence of mirror reflection symmetry $\bss$, each eigenmode can be labeled by their $\bss$ eigenvalue $q_{E}(\bss)$:
\bea\label{mirror:single-spinon mode}
&\hat U_\bss\gamma_E\hat U_\bss^{-1}=q_E(\bss)\gamma_E,\\
&[q_E(\bss)]^2=\omega_\bss=\pm1.\notag
\eea
On the other hand, mirror reflection $R_x$ satisfying (\ref{criterion 1: two mirrors}) leads to another degenerate single-spinon level $\hat U_{R_x}\gamma_E\hat U_{R_x}^{-1}$ with an opposite $\bss$ eigenvalue $-q_E(\bss)$ since
\bea
\notag&\hat U_\bss\big(\hat U_{R_x}\gamma_E\hat U_{R_x}^{-1}\big)\hat U_\bss^{-1}=-\hat U_{R_x}\hat U_\bss\gamma_E\hat U_\bss^{-1}\hat U_{R_x}^{-1}\\
&=-q_E(\bss)\cdot\big(\hat U_{R_x}\gamma_E\hat U_{R_x}^{-1}\big)
\eea
Therefore each single-spinon level must be at least 2-fold degenerate, due to anti-commuting mirror operations $\hat U_\bss$ and $\hat U_{R_x}$ on spinons as shown in (\ref{criterion 1: two mirrors}).

Now consider a lattice with $N$=~odd number of lattice sites with one half-integer spin-$S$ on each site, such as a odd by odd square lattice on FIG. \ref{fig:square}. Since the number of both positive and negative energy eigenstates is $2NS$=~odd, and each energy level is at least two-fold degenerate, there must be (at least) two degenerate zero-energy eigenstates in the single-spinon spectrum. Therefore at mean-field level, the spinon/parton ground state must have at least 2-fold degeneracy.

The above argument can be easily generalized to interacting parton/spinon Hamiltonian (\ref{nambu basis}) on an $N$=~odd lattice. Since in the Nambu basis, the many-body ground state $\dket{\Psi_{\text{parton}}}$ has a conserved particle number ${\hat F}=2NS$=~odd, two mirror reflections $\hat U_\bss$ and $\hat U_{R_x}$ anticommute on the many-body ground state as shown in (\ref{proj symm:nambu spinon}), leading to at least 2-fold degeneracy. However, there are some subtlety in defining a \ztsl~with $N$=odd half-integer spins, where the physical many-spin ground state contains an odd number of spinons and must be at least 2-fold degenerate due to Kramers theorem. Below we put previous intuitive arguments on a firmer ground, by using Schmidt decomposition of the many-spinon ground state on an infinite cylinder\cite{Watanabe2015}.

Consider interacting spinon Hamiltonian (\ref{nambu basis}) on an infinite cylinder wrapped along $\hat y$ direction in FIG. \ref{fig:square}, where the cylinder circumference is $L_y$=odd. Assume it hosts a unique gapped ground state $\dket{\Psi_{\text{parton}}}$, preserving mirror reflections $R_x,\bss$ and time reversal symmetry $\bst$. This gapped ground state is also a short-range-entangled (SRE) state\cite{Chen2011a} on this infinite cylinder. Across the $\bss$-preserving entanglement cut denoted by the red dotted-dash line in FIG. \ref{fig:square}, the SRE ground state can be Schmidt decomposed into
\bea
\spgs=\sum_{\nu}w_{\nu}~\dket{\nu}_{L}\otimes\dket{\nu}_{\bar L}
\eea
where $w_{\nu}$ are Schmidt weights. $\{\dket{\nu}_{L}\}$ and $\{\dket{\nu}_{\bar L}\}$ denote a set of orthonormal basis for the Hilbert space of region $L$ and $\bar L$. The Schmidt spectrum $\{w_{\nu}\}$ is discrete for SRE state $\spgs$ . Since the bipartition (red line) of the lattice preserves mirror $\bss$, the orthonormal basis can be labeled by their reflection quantum number $q(\bss)$:
\bea
\hat U_\bss\dket{\nu}_{L}=q_{\nu,L}(\bss)\dket{\nu}_{L},~~\hat U_\bss\dket{\nu}_{\bar L}=q_{\nu,\bar L}(\bss)\dket{\nu}_{\bar L}.
\eea
which must satisfy
\bea\label{total mirror eigenvalue}
q_{\nu,L}(\bss)q_{\nu,\bar L}(\bss)\equiv q_\bss=\pm1,~~~\forall~\nu.
\eea
so that parton/spinon ground state $\spgs$ preserves mirror $\bss$ with eigenvalue $q_\bss$.

Now let's consider another entanglement cut related to the previous one by mirror reflection $R_x$, depicted by the green dotted-dash line in FIG. \ref{fig:square}. The Schmidt decomposition across this cut is given by
\bea
\spgs=\sum_{\nu}w_\nu^\prime\dket{\nu}_R\otimes\dket{\nu}_{\bar R},
\eea
where by mirror reflection we can choose
\bea
w_\nu^\prime=w_\nu,~~\dket{\nu}_R=\hat U_{R_x}\dket{\nu}_L,~~\dket{\nu}_{\bar R}=\hat U_{R_x}\dket{\nu}_{\bar L}.
\eea
Note that on an infinite cylinder, the total particle number $\hat F_{L}=\sum_{{\bf r}\in L}\psi^\dagger_{\bf r}\psi_{\bf r}$ on the l.h.s. of an entanglement cut is ill-defined in the thermodynamic limit\cite{Watanabe2015}, however the charge fluctuation around its average value
\bea\label{number fluctuation}
\hat Q_L\equiv\sum_{{\bf r}\in L}(\psi^\dagger_{\bf r}\psi_{\bf r}-2S)=-\hat Q_{\bar L}\in\mbz,
\eea
is still well-defined. Hence we have
\bea
\notag&\notag(-1)^{\hat F_L}\dket{\nu}_{L}=e^{\imth\Phi}(-1)^{Q_{\nu,L}}\dket{\nu}_{L},~~\forall~\nu,\\
&\notag\hat Q_L\dket{\nu}_{L}=Q_{\nu,L}\dket{\nu}_{L},\\
&\notag(-1)^{\hat F_{\bar L}}\dket{\nu}_{\bar L}=e^{-\imth\Phi}(-1)^{Q_{\nu,\bar L}}\dket{\nu}_{\bar L},~~\forall~\nu,\\
&\hat Q_{\bar L}\dket{\nu}_{\bar L}=Q_{\nu,\bar L}\dket{\nu}_{\bar L}.\label{number parity}
\eea
where $\Phi$ is an ambiguous but $\nu$-independent phase\cite{Watanabe2015}. By mirror symmetry $R_x$ it's also clear that
\bea\label{mirror:number fluctuation}
Q_{\nu,R}=Q_{\nu,L}=-Q_{\nu,\bar L}=-Q_{\nu,\bar R}\in\mbz
\eea
Making use of symmetry action (\ref{criterion 1: two mirrors}), it's straightforward to work out the mirror $\bss$-eigenvalues of Schmidt components as
\bea
\notag&\hat U_\bss\dket{\nu}_{R}=e^{\imth\Phi}(-1)^{Q_{\nu,L}}q_{\nu,L}(\bss)\dket{\nu}_{R},\\
&\hat U_\bss\dket{\nu}_{\bar R}=e^{-\imth\Phi}(-1)^{Q_{\nu,\bar L}}q_{\nu,\bar L}(\bss)\dket{\nu}_{\bar R}.
\eea
Applying mirror reflection $\bss$ twice and requiring $q_{\nu,L}(\bss^2)=q_{\nu,R}(\bss^2)=1$, we obtain that\cite{Pollmann2010}
\bea\label{ambiguous phase}
e^{2\imth\Phi}=1.
\eea
On the other hand notice that region $\bar L=M\bigcup R$ (see FIG. \ref{fig:square}), hence the Schmidt components on the r.h.s. of the two cuts are related by addition of region $M$ (a column containing $L_y$=odd sites):
\bea\label{Schmidt comp:bar L}
\dket{\nu}_{\bar L}=\sum_{\mu,\rho}B^\rho_{\mu\nu}\dket{\rho}_M\otimes\dket{\mu}_R,
\eea
where $\{\dket{\rho}_M\}$ is a set of orthonormal basis for Hilbert space in region $M$, with particle number satisfying
\bea
&\hat F_{M}\dket{\rho}_M=(Q_{\rho,M}+L_yS)\dket{\rho}_M,\\
&Q_{\nu,\bar L}=Q_{\rho,M}+Q_{\mu,R},~~~\forall~B^{\rho}_{\mu\nu}\neq0.
\eea
Now let's perform a series of symmetry operations $\hat U_\bss\hat U_{R_x}\hat U_\bss^{-1}\hat U_{R_x}^{-1}$ on both sides of state (\ref{Schmidt comp:bar L}), and we reach a contradiction to (\ref{ambiguous phase})
\bea
e^{\imth\Phi}=(-1)^{L_yS}e^{-\imth\Phi}
\eea
for a $L_y$=~odd cylinder. Therefore we've shown that symmetry implementation (\ref{criterion 1: two mirrors}) is incompatible with a unique gapped symmetric ground state $\spgs$ of any interacting parton/spinon Hamiltonian (\ref{nambu basis}). Since the above arguments is valid for any odd circumference $L_y=1\mod2$ of the cylinder, it is naturally applicable to a two-dimensional system.

As a result, we have established that for spinon symmetry fractionalization class (\ref{criterion 1: two mirrors}), with two perpendicular mirror planes, time reversal and $U(1)_{{\bf S}^z}$ spin rotational symmetries, it is impossible to have a gapped symmetric \ztsl. This also proves the stability of gapless symmetric \ztsl s protected by spinon fractionalization class (\ref{criterion 1: two mirrors}).

\subsection{Mirror plane crossing an odd number of half-integer spins}

The 2nd criterion for gapless $Z_2$ spin liquids is the following implementation of time reversal $\bst$ and a mirror reflection (say, $R_x$ in FIG. \ref{fig:square}), whose mirror planes crosses an odd number of half-integer spins (\ie odd sites):
\bea\notag
&\bst R_x\bst^{-1}R_x=\bse\Longrightarrow\\
&\label{criterion 2:mirror+TRS}
{\omega_{R_x\bst}}{\omega_{R_x}}=-1\Leftrightarrow\hat\bst\hat U_{R_x}\hat\bst^{-1}\hat U_{R_x}=(-1)^{\hat F}.
\eea
This gauge-invariant phase factor of symmetry fractionalization can also be written as
\bea\notag
\frac{\omega(\bst R_x,\bst R_x)}{\omega(\bst,\bst)}=\frac{\omega(\bst,R_x)}{\omega(R_x,\bst)}\cdot\omega(R_x,R_x)={\omega_{R_x\bst}}{\omega_{R_x}}=-1
\eea

First at single-particle level \ie for the parton mean-field Ansatz (\ref{mean-field ham diagonalized}), we can label single-spinon modes by their mirror eigenvalues $q_E(R_x)$ as defined in (\ref{mirror:single-spinon mode}). Performing time reversal operation on each mode will lead to another mode with the opposite energy
\bea
&\notag\bst\gamma_E\bst^{-1}=\gamma_{-E}^\dagger,~~~|q_E(R_x)|^2=1\Longrightarrow\\
\notag&\hat U_{R_x}\gamma_{-E}^\dagger\hat U_{R_x}^{-1}=-\bst\hat U_{R_x}^{-1}\gamma_E\hat U_{R_x}\bst^{-1}=-q_E(R_x)\gamma_{-E}^\dagger\\
&\Longrightarrow\hat U_{R_x}\gamma_{-E}\hat U_{R_x}^{-1}=-q^\ast_E(R_x)\gamma_{-E}.
\eea
Let's first assume there is a finite energy gap in the single-spinon spectrum, separating positive energy states from those with negative energies. Now that each positive energy level $\gamma_E$ always shows up in pairs with its time-reversal partner $\gamma_{-E}$, the total mirror eigenvalue of all energy levels is given by
\bea
&\notag\hat U_{R_x}\big(\prod_{E}\gamma_E\big)\hat U_{R_x}^{-1}=\\
&\prod_{E>0}\Big(q_E(R_x)\cdot\big[-q_E^\ast(R_x)\big]\Big)=(-1)^{2NS}.
\eea
This leads to a total mirror eigenvalue $-1$ of all energy levels on a lattice with $N=$~odd sites, which is impossible in any physical parton/spinon Hamiltonian. This contradiction can only be resolved by the existence of zero-energy single-spinon modes, \ie a gapless spinon spectrum.

Below we prove that a unique SRE symmetric ground state of interacting spinon/parton Hamiltonian (\ref{nambu basis}) is impossible on an infinite $R_x$-preserving cylinder with circumference $L_y$=odd (see FIG. \ref{fig:square}), again using its Schmidt decomposition. Assuming a unique SRE ground state, again it can be Schmidt decomposed into
\bea
&\notag\spgs=\sum_{\nu}w_\nu\dket{\nu}_L\otimes\dket{\nu}_{\bar L}
=\sum_\nu w_\nu\dket{\nu}_R\otimes\dket{\nu}_{\bar R},\\
&\dket{\nu}_R=\hat U_{R_x}\dket{\nu}_L,~~~\dket{\nu}_{\bar R}=\hat U_{R_x}\dket{\nu}_{\bar L}.
\eea
Time reversal operation (\ref{nambu:time reversal}) in the Nambu basis indicates that
\bea
\hat\bst^2\dket{\nu}_L=(-1)^{\sum_{{\bf r}\in L}(\psi^\dagger_{\bf r}\psi_{\bf r}-2S)}\dket{\nu}_L=(-1)^{Q_{\nu,L}}\dket{\nu}_L.
\eea
where $Q_{\nu,L}$ are particle number fluctuation in region $L$, defined in (\ref{number fluctuation})-(\ref{mirror:number fluctuation}). Again since the parton density $2S\in\mbz$ is an integer in the Nambu basis, we have the following relation
\bea
e^{2\imth\Phi}=1
\eea
for ambiguous phase $\Phi$ in (\ref{number parity}) for the Nambu particle parity of Schmidt eigenstates. Due to mirror reflection symmetry $R_x$ we have
\bea
\notag&\hat\bst\hat U_{R_x}\hat\bst^{-1}\hat U_{R_x}\dket{\nu}_R=(-1)^{\hat F_R}\dket{\nu}_R=e^{\imth\Phi}(-1)^{Q_{\nu,R}}\dket{\nu}_R,\\
&\hat\bst\hat U_{R_x}\hat\bst^{-1}\hat U_{R_x}\dket{\nu}_{\bar L}=e^{-\imth\Phi}(-1)^{Q_{\nu,\bar L}}\dket{\nu}_{\bar L}.
\eea
Following the same strategy ($\bar L=M\bigcup R$) we used in proving the 1st criterion, by acting $\hat\bst\hat U_{R_x}\hat\bst^{-1}\hat U_{R_x}$ on both sides of (\ref{Schmidt comp:bar L}) we end up with the following contradiction
\bea
e^{-\imth\Phi}=(-1)^{L_yS}e^{\imth\Phi}
\eea
on a cylinder with circumference $L_y$=odd.

Therefore we've proved that symmetry fractionalization class (\ref{criterion 2:mirror+TRS}) of spinons is incompatible with a symmetric SRE ground state for the generic interacting parton/spinon Hamiltonian (\ref{nambu basis}). This therefore leads to stable gapless $Z_2$ spin liquids protected by mirror and time reversal symmetries, characterized by spinon symmetry fractionalization (\ref{criterion 2:mirror+TRS}).

\subsection{$C_{2n}$ rotation or inversion centered at a half-integer spin}

The 3rd criterion is provided by the interplay of $2n$-fold site-centered rotation $C_{2n}$ and time reversal operation $\bst$. In particular, a $C_{2n}$ rotation necessarily imply a two-fold rotation
\bea\label{C2n->C2_inversion}
\hat C_2=(\hat C_{2n})^n\equiv\hat I.
\eea
around the same site with an half-integer spin on it. This 2-fold rotation can also be viewed as the inversion operation $\hat I$ for the 2d plane. Our 3rd criterion for gapless \ztsl s is guaranteed by the following spinon symmetry fractionalization class
\bea\notag
&\bst C_2\bst^{-1}C_2=\bse\Longrightarrow\\
&\label{criterion 3:C2+TRS}
\frac{\omega(C_2\bst,C_2\bst)}{\omega(\bst,\bst)}=-1\Leftrightarrow\hat\bst\hat U_{C_2}\hat\bst^{-1}\hat U_{C_2}=(-1)^{\hat F}.
\eea
The above symmetry fractionalization class is quite analogous to previous criterion (\ref{criterion 2:mirror+TRS}) for reflection $\hat R_x$ and time reversal $\bst$, where the only difference is to replace mirror $\hat R_x$ by inversion $\hat C_2=\hat I$. It's straightforward to show that our previous arguments on criterion (\ref{criterion 2:mirror+TRS}) can be applied completely in parallel to prove the above criterion (\ref{criterion 3:C2+TRS}).

The above criterion can be further refined for $2n$-fold rotation $C_{2n}$, depending on whether $n$ is an even or odd integer\footnote{A crucial difference between an even-rank rotation $C_{2n}$ and odd-rank rotation $C_{2n+1}$ is that two-dimensional inversion $\hat I=\hat C_2$ can be generated by even-rank rotation $I=(C_{2n})^n$ but not the odd-rank rotation. Therefore our 3rd criterion only applies to even-rank rotation $C_{2n}$.}. It will lead to two different criteria for gapless \ztsl s protected by $C_{2n}$ rotations, as we elaborate below.

First of all, due to algebraic equality (\ref{C2n->C2_inversion}), it's straightforward to verify the following relations for gauge-invariant phase factors of spinon symmetry fractionalization:
\bea
&\omega(C_2,C_2)=\prod_{i=1}^{2n-1}\omega\big(C_{2n},(C_{2n})^i\big)\equiv\omega_{C_{2n}},\\
&\frac{\omega(C_2,\bst)}{\omega(\bst,C_2)}=\big[\frac{\omega(C_{2n},\bst)}{\omega(\bst,C_{2n})}\big]^n
\eea
Meanwhile criterion (\ref{criterion 3:C2+TRS}) is the same as
\bea
\frac{\omega(C_2\bst,C_2\bst)}{\omega(\bst,\bst)}=\frac{\omega(C_2,\bst)}{\omega(\bst,C_2)}\omega(C_2,C_2)=-1
\eea
Note that $\omega(g,h)=\pm1$ as elements of 2nd group cohomology $\mathcal{H}^2(G_s,\mbz_2)$, and therefore criterion (\ref{criterion 3:C2+TRS}) can be rewritten as follows:
\bea\notag
&\text{if}~n=\text{odd}:~~~~~~\frac{\omega(C_{2n},\bst)}{\omega(\bst,C_{2n})}=-\omega_{C_{2n}}\Leftrightarrow\\
\label{criterion 3:n=odd}&\hat U_{C_{2n}}\bst\hat U_{C_{2n}}^{-1}\bst^{-1}=(-1)^{\hat F}\cdot \big(\hat U_{C_{cn}})^{2n}.
\eea
and
\bea
\notag&\text{if}~n=\text{even}:~~~~~~\omega_{C_{2n}}=-1\Leftrightarrow\\
\label{criterion 3:n=even}&\big(\hat U_{C_{cn}})^{2n}=(-1)^{\hat F}.
\eea

In summary, in the presence of a 2-fold rotational symmetry $C_2$ centered at a half-integer spin, the 3rd criterion dictates that spinon symmetry fractionalization (\ref{criterion 3:C2+TRS}) is incompatible with any gapped symmetric \ztsl. It also establishes the stability of gapless \ztsl s protected by such a $C_2$ rotational symmetry.

Now that a $2n$-fold crystalline rotation $C_{2n}$ necessarily leads to a 2-fold rotation as shown in (\ref{C2n->C2_inversion}), our 3rd criterion (\ref{criterion 3:C2+TRS}) further leads to two criteria (\ref{criterion 3:n=odd})-(\ref{criterion 3:n=even}) for stable gapless \ztsl s, protected by a $2n$-fold rotation $C_{2n}$ centered at a half-integer spin. Specifically criterion (\ref{criterion 3:n=odd}) applies to $n=$~odd case, while criterion (\ref{criterion 3:n=even}) applies to $n=$~even case.

\begin{table}[tb!]
\begin{tabular} {|c|c|c|c|}
\hline
Algebraic Identity&\multirow{2}{1.3cm}{SB $b_{\alpha}$ in \Ref{Wang2006}} &
\multirow{2}{1.3cm}{AF $f_{\alpha}$ in \Ref{Lu2016}}&\multirow{2}{2.3cm}{fractionalization class of spinons}\\
&&&\\ \hline
$T^{-1}_{2}T^{-1}_{1}T_{2}T_{1}=\bse$&(-1)$^{p_1}$&$\eta_{12}$&$\omega_{T_1T_2}$\\ \hline
$\bss^{-1}T_1\bss T_2^{-1}=\bse$&1& 1&1\\ \hline
$\bss^{-1}T_2\bss T_1^{-1}=\bse$&1&1&1\\ \hline
$\cs^{-1}T_1\cs T_2=\bse$&1&1&1\\ \hline
$\cs^{-1}T_2\cs T_2^{-1}T_1^{-1}=\bse$&1&1&1\\ \hline
$\bss^2=\bse$&(-1)$^{p_2}$&$\eta_\bss$&$\omega_\bss$\\ \hline
$R^{2}=(\cs\bss)^2=\bse$&(-1)$^{p_2+p_3}$&$\eta_{\bss\cs}$&$\omega_{R}$\\ \hline
$(\cs)^{6}=\bse$&(-1)$^{p_3}$&$\eta_\cs$&$\omega_{\cs}$\\ \hline
$T_1^{-1}\bst^{-1}T_1\bst=\bse$&1&1&1\\ \hline
$T_2^{-1}\bst^{-1}T_2\bst=\bse$&1&1&1\\ \hline
$\bss^{-1}\bst^{-1}\bss\bst=\bse$&(-1)$^{p_2}$&$\eta_{\bss\bst}$&$\omega_{\bss\bst}$\\ \hline
$R^{-1}\bst^{-1}R \bst=\bse$&(-1)$^{p_2+p_3}$&$\eta_{\cs\bst}\eta_{\bss\bst}$&$\omega_{R\bst}$\\ \hline
$\bst^2=\bse$&-1&-1&-1\\ \hline
\end{tabular}
\caption{Triangular lattice: symmetry fractionalization class $(\mbz_2)^6\subset\mathcal{H}^2(P6mm\times Z_2^\bst,\mbz_2)$ of spinons in half-integer-spin $Z_2$ spin liquids, and their realizations in $S=\frac12$ Schwinger-boson (SB) and Abrikosov-fermion (AF) representations, following the convention of \Ref{Wang2006,Lu2016}. The fractionalization classes from 2nd group cohomology\cite{Zheng2015} are labeled by six $Z_2$-valued integers $\omega=\pm1$. We show that a $Z_2$ spin liquid is gapless if violating any of the following conditions: $\omega_{\bss}=\omega_{\bss\bst}$, $\omega_{R}=\omega_{R\bst}$ and $\omega_{\cs}=\omega_{\bss}\omega_{R}$. This leads to only $2^3=8$ distinct gapped $Z_2$ spin liquids, all realized in both SB \cite{Sachdev1992,Wang2006} and AF representations\cite{Lu2016}. There are 12 symmetry protected gapless \ztsl s in the AF representation.}
\label{tab:triangle}
\end{table}

\begin{figure}
\includegraphics[width=0.9\columnwidth]{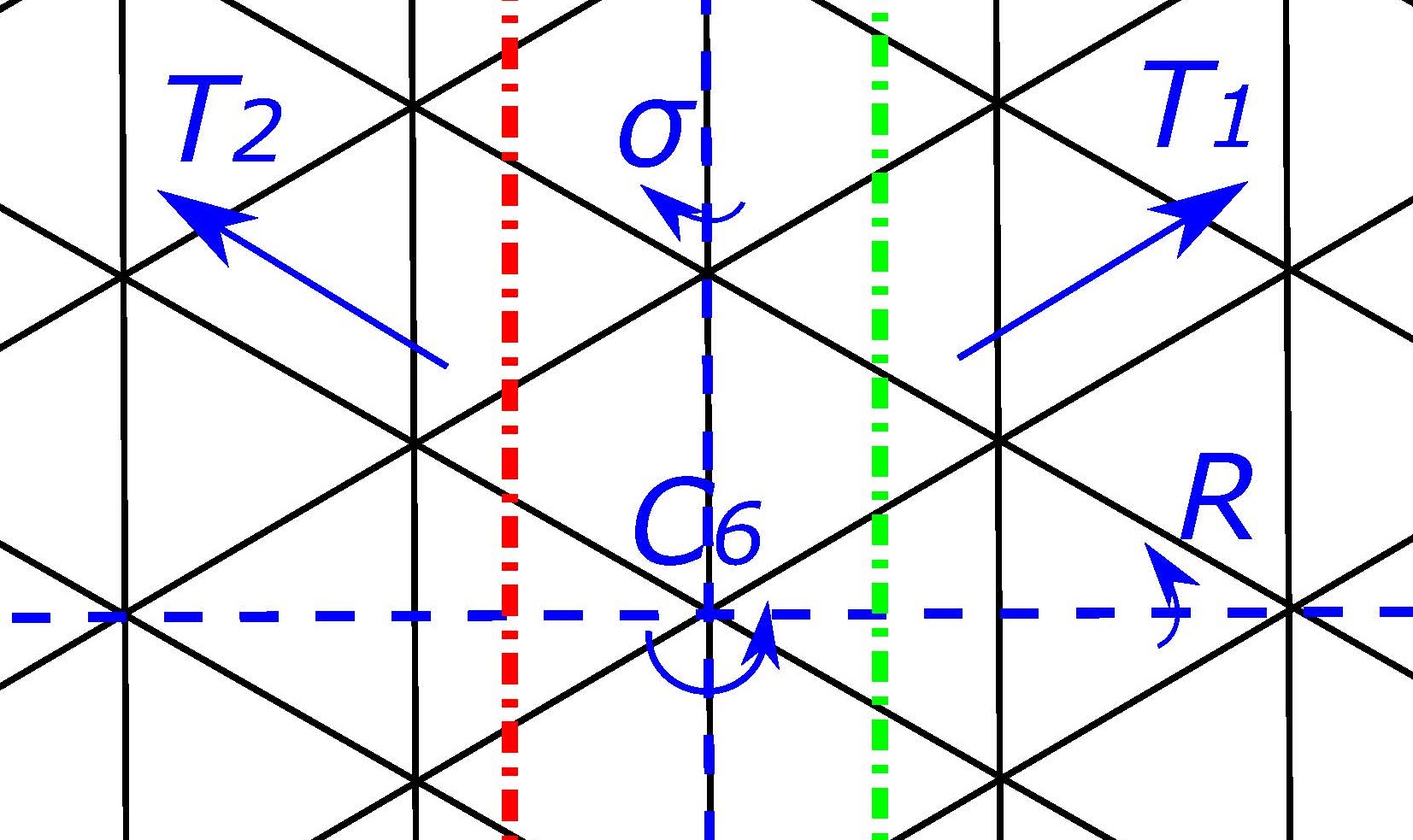}
\caption{(color online) Space group ($P6mm$) symmetries of triangular lattice, generated by two translations $T_{1,2}$, mirror reflection $\bss$ and site-centered 6-fold rotation $\cs$. Another mirror reflection is introduced as $R=(\cs)^3\bss$. The two mirror planes ($\bss$ and $R$) intersect at a lattice site. Red and green dotted dash lines denote two entanglement cuts of the lattice, related by mirror $\bss$ or inversion $I=(\cs)^3$. }
\label{fig:triangle}
\end{figure}

In the following we apply the above three criteria (\ref{criterion 1: two mirrors}), (\ref{criterion 2:mirror+TRS}) and (\ref{criterion 3:C2+TRS}) to a few examples. These are \ztsl s of spin-$1/2$ systems on square, triangular and kagome lattices with $SU(2)$ spin rotation (it contains $U(1)_{{\bf S}^z}$ spin rotations as a subgroup), time reversal and space group symmetries. Among all classified symmetric \ztsl s in the Abrikosov-fermion representation, where $2S=1$ and partons $\{\phi_{{\bf r},\alpha}\equiv f_{{\bf r},\alpha}\}$ are fermions, we will show that a large part of them are gapless \ztsl s protected by certain crystalline symmetries. All symmetric gapped \ztsl s in the Abrikosov-fermion representation have their counterparts in the Schwinger-boson representation, where $2S=1$ and $\{\phi_{{\bf r},\alpha}\equiv b_{{\bf r},\alpha}\}$ are hard-core bosons.

\section{Gapped and gapless $Z_2$ spin liquids on the square lattice}\label{SQUARE}

The space group of square lattice is $P4gm$, generated by two translations $T_{1,2}$, mirror reflection $\bss$ and 4-fold rotation $C_4$ as shown in FIG. \ref{fig:square}. The symmetry implementations on bosonic/fermionic spinons of symmetric $Z_2$ spin liquids on the square lattice are summarized in TABLE \ref{tab:square}. The last column of the table lists the algebraic symmetry fractionalization classes $(\mbz_2)^9\subset\mathcal{H}^2(P4gm\times Z_2^\bst,\mbz_2)$ of spinons, as computed in \Ref{Essin2013}. Since in a half-integer spin system, spinons carry half-integer spins and are Kramers doublets of time reversal symmetry satisfying $\hat\bst^2=(-1)^{\hat F}$, each spinon symmetry fractionalization class is labeled by 9 $Z_2$-valued integers $\{\omega=\pm1\}$\cite{Essin2013}.

In spin-$1/2$ case ($2S=1$), Schwinger-boson (SB) and Abrikosov-fermion (AF) representations provide two concrete constructions for symmetric \ztsl s associated with the $2^9=512$ symmetry fractionalization classes. In the SB representation\cite{Schwinger1965}, the partons $\{\phi_{{\bf r},\alpha}\equiv b_{{\bf r},\alpha}\}$ are hard-core bosons, while the partons $\{\phi_{{\bf r},\alpha}\equiv f_{{\bf r},\alpha}\}$ become fermions in the AF representation\cite{Abrikosov1965}. A detailed classification\cite{Yang2016} within the SB construction leads to $2^6=32$ different states labeled by 6 $Z_2$-valued integers $\{p_i=0,1|1i=1,2,3,4,7,8\}$; while in AF construction one can write down 176=$11\times2^4$ different \ztsl~states\cite{Wen2002}. In these SB states, the gauge-invariant phase factors are given by 6 $Z_2$-valued phases $\{(-1)^{p_i}=\pm1\}$, while in AF states they are given by 9 $Z_2$-valued phases $\{\eta=\pm1\}$, as summarized in TABLE \ref{tab:square}. For instance, the gauge-invariant phase factor $\omega_{T_1T_2}$ associated with algebraic identity $T_2^{-1}T_1^{-1}T_2T_1=\bse$ is equal to $(-1)^{p_1}$ when acting on Schwinger bosons, or $\eta_{xy}=\pm1$ when acting on Abrikosov fermions. These symmetry implementations on bosonic/fermionic spinons $b_{\alpha}/f_\alpha$ in the SB/AF representation, as well as their correspondence to the algebraic spinon symmetry fractionalization class are all listed in TABLE \ref{tab:square}.

Based on our previous analysis, in the general parton construction (\ref{general parton construction}) of half-integer-spin systems, the spinon symmetry fractionalization class satisfying criteria (\ref{criterion 1: two mirrors}), (\ref{criterion 2:mirror+TRS}), (\ref{criterion 3:C2+TRS}) and (\ref{criterion 3:n=odd})-(\ref{criterion 3:n=even}) will necessarily lead to gapless $Z_2$ spin liquids. On the square lattice, there are two perpendicular mirror planes $\bss$ and $R_x=C_4\bss (C_4)^{-1}$ in FIG. \ref{fig:square}, that intersects at one lattice site with a spin-$\frac12$. Since the two mirror planes are related by a 90 degree rotation $C_4$, they share the same gauge-invariant phase factor of spinon symmetry fractionalization
\bea
\omega(\bss,\bss)=\omega(R_x,R_x)\equiv\omega_\bss.
\eea
The combination of the two mirror is nothing but the 2-fold rotation (or inversion) around a lattice site:
\bea
C_2=(C_4)^2=R_x\bss.
\eea
and we have
\bea
\omega(C_2,C_2)=\omega(C_4,C_4)\omega\big(C_4,C_2)\omega\big(C_4,(C_4)^3\big)\equiv\omega_{C_4}.
\eea
Therefore the 1st gapless criterion (\ref{criterion 1: two mirrors}) can be written as
\bea\label{square:gapless:two mirror}
\frac{\omega(\bss,R_x)}{\omega(R_x,\bss)}=\frac{\omega(C_2,C_2)}{\omega(\bss,\bss)\omega(R_x,R_x)}=\omega_{C_4}=-1.
\eea
on square lattice.

For 2nd gapless criterion (\ref{criterion 2:mirror+TRS}), its application to mirror plane $\bss$ leads to
\bea\label{square:gapless:bss}
\omega_\bss\omega_{\bss\bst}=-1.
\eea
Meanwhile, applying the same criterion (\ref{criterion 2:mirror+TRS}) to another mirror plane $R_{xy}=C_4\bss$ leads to
\bea\label{square:gapless:Rxy}
\omega_R\omega_{R\bst}=-1.
\eea

Finally for 4-fold rotational symmetry $C_4$, applying the 3rd gapless criterion (\ref{criterion 3:n=even}) also leads to condition (\ref{square:gapless:two mirror}).

To summarize, if any of the above 3 conditions (\ref{square:gapless:two mirror})-(\ref{square:gapless:Rxy}) are satisfied for the spinon fractionalization class, we will end up with a symmetry protected gapless \ztsl~robust against any symmetry-preserving perturbations. As a result, any gapped symmetric \ztsl~on square lattice must obey the following rules of spinon symmetry fractionalization:
\bea\label{square:gap condition}
\omega_\cf=\omega_{\bss}\omega_{\bss\bst}=\omega_{R}\omega_{R\bst}=1.
\eea
This ``gap condition'' applies to both bosonic and fermionic spinons.

From the 2nd column of TABLE \ref{tab:square}, it's straightforward to see that the above condition (\ref{square:gap condition}) is automatically satisfied for all Schwinger boson mean-field states of \ztsl s on square lattice, since
\bea
&\notag\omega_\cf^b=1,~~\omega_\bss^b=\omega_{\bss\bst}^b=(-1)^{p_4},\\
&\omega_R^b=\omega_{R\bst}^b=(-1)^{p_4+p_7}.
\eea
for the symmetry fractionalization class of bosonic spinons $\{b_{{\bf r},\alpha}\}$ in the Schwinger-boson (SB) representation. This means all 64 Schwinger boson mean-field states are generically gapped \ztsl s\cite{Yang2016}.

On the other hand, in the Abrikosov-fermion construction\cite{Wen2002} (3rd column of TABLE \ref{tab:square}) of symmetric \ztsl s on square lattice, the ``gap condition'' (\ref{square:gap condition}) dictates that a gapped \ztsl~state must satisfy
\bea
\eta_{\cf}=\eta_\bss\eta_{\bss\bst}=\eta_{\bss\cf}\eta_{\cf\bst}=1.
\eea
since the symmetry fractionalization class of fermionic spinons $\{f_{{\bf r},\alpha}\}$ are given by
\bea
\notag&\omega^f_\cf=\eta_\cf,~~\omega^f_\bss=\eta_\bss,~~\omega^f_{\bss\bst}=\eta_{\bss\bst},\\
&\omega^f_R=\eta_\bss\eta_{\bss\cf},~~\omega^f_{R\bst}=\eta_{\cf\bst}\eta_{\bss\bst}.
\eea
in the Abrikosov-fermion (AF) representation\cite{Wen2002}. Among all AF mean-field states, again only $2^6=64$ states satisfy the above gap condition and lead to gapped symmetric \ztsl s.

One natural question is: what is the relation between the 64 \ztsl s in SB representation and the 4 states in AF representation? In fact, it is not coincidental to have 64 gapped symmetric \ztsl s in both SB and AF representations: there is a one-to-one correspondence between the 64 SB states and the 64 AF states. A gapped \ztsl~is topologically ordered\cite{Wen2004B} and hosts 3 types of gapped anyon excitations: bosonic spinon $b$, fermionic spinon $f$ and bosonic vison $v$ obeying the following fusion rule
\bea\label{fusion rule}
f=b\times v
\eea
Consequently, the phase factors $\{\omega^a(g,h)|g,h\in G_s;a=b,f,v\}$ characterizing the symmetry fractionalization class of anyons must satisfy the following relation:
\bea\label{fusion rule:cohomology}
\omega^f(g,h)=\lambda(g,h)\cdot\omega^v(g,h)\cdot\omega^b(g,h)
\eea
where $\lambda(g,h)=\pm1$ are extra twisting factors\cite{Essin2013,Qi2015,Zaletel2017,Lu2017} for symmetry operations $g,h\in G_s$. As shown in \Ref{Qi2015a,Zaletel2017,Lu2017}, for any gapped symmetric \ztsl~on the square lattice, the vison fractionalization class $\{\omega^v(g,h)\}$ can be uniquely determined. Therefore relation (\ref{fusion rule:cohomology}) establishes the one-to-one correspondence between bosonic spinon fractionalization class $\{\omega^b(g,h)\}$ and the fermionic spinon one $\{\omega^f(g,h)\}$. In particular, the bosonic partons in SB representation are the bosonic spinons $b$, while fermionic partons in AF representation are the fermionic spinons $f$. Therefore we can establish a duality between the gapped \ztsl s in SB representation and those in AF representation. On the square lattice as shown in \Ref{Yang2016}, this duality map is given by the following identities:
\bea
&\notag\eta_{xy}=(-1)^{p_1+1},~~~\eta_{xpx}=(-1)^{p_3},~~~\eta_{xpy}=(-1)^{p_2+1},\\
&\notag\eta_\bss=\eta_{\bss\bst}=(-1)^{p_4+1},~~~\eta_{\cf\bst}=\eta_{\bss\cf}=(-1)^{p_7},\\
&\eta_t=(-1)^{p_8},~~~\eta_\cf=1.
\eea
All $64=4\times2^4$ gapped \ztsl s can be realized by mean-field Ansatz in both SB and AF representations. More concretely in the AF representation, the fermionic partons in all $64=2^4\times 4$ gapped \ztsl s have the following space group symmetry implementations
\bea
&\notag\eta_{xy},\eta_{xpx},\eta_{xpy},\eta_t=\pm1,\\
&\notag(g_{P_{xy}},g_{P_x},g_{P_y},g_\bst)=(\tau^0,\tau^0,\tau^0,\imth\tau^3),\\
&\notag(g_{P_{xy}},g_{P_x},g_{P_y},g_\bst)=(\tau^0,\imth\tau^1,\imth\tau^1,\imth\tau^3),\\
&\notag(g_{P_{xy}},g_{P_x},g_{P_y},g_\bst)=(\imth\tau^1,\tau^0,\tau^0,\imth\tau^3),\\
&\notag(g_{P_{xy}},g_{P_x},g_{P_y},g_\bst)=(\imth\tau^1,\imth\tau^1,\imth\tau^1,\imth\tau^3).
\eea
in the notation of \Ref{Wen2002}.\\

While all SB mean-field states are gapped \ztsl s, many states in the AF representation\cite{Wen2002,Wen2002a} correspond to symmetry-protected gapless \ztsl s. In particular, in total there are $160=10\times2^4$ AF mean-field states where fermionic spinons $\{f_{{\bf r},\alpha}\}$ are Kramers doublets of time reversal symmetry with $\bst^2=(-1)^{\hat F}$. Besides the 64 gapped \ztsl s, the remaining $6\times2^4=96$ AF states with Kramers-doublet spinons all satisfy at least one of the 3 gapless conditions (\ref{square:gapless:two mirror})-(\ref{square:gapless:Rxy}). Therefore as we discussed, these 96 states in \Ref{Wen2002} all realize symmetry-protected gapless \ztsl s, which are robust against any symmetry-preserving perturbations.

\begin{table}[tb!]
\begin{tabular} {|c|c|c|c|}
\hline
Algebraic Identity&\multirow{2}{1.3cm}{SB $b_{\alpha}$ in \Ref{Wang2006}} &
\multirow{2}{1.3cm}{AF $f_{\alpha}$ in \Ref{Lu2011}}&\multirow{2}{2.3cm}{fractionalization class of spinons}\\
&&&\\ \hline
$T^{-1}_{2}T^{-1}_{1}T_{2}T_{1}$&(-1)$^{p_1}$&$\eta_{12}$&$\omega_{T_1T_2}$\\ \hline
$T^{-1}_{1}\cs^{-1}T_{2}\cs$&1& 1&1\\ \hline
$T^{-1}_{1}T_{2}\cs^{-1}T_{1}\cs$&1&1&1\\ \hline
$T^{-1}_{1}R^{-1}_{y}T_{1}R_y$&(-1)$^{p_1}$&$\eta_{12}$&$\omega_{T_1T_2}$\\ \hline
$T^{-1}_{1}T_{2}R^{-1}_{y}T_{2}R_{y}$&(-1)$^{p_1}$&$\eta_{12}$&$\omega_{T_1T_2}$\\ \hline
$R_x^2=(\cs R_{y})^2$&(-1)$^{p_2+p_3}$&$\eta_\bss$&$\omega_{R_x}$\\ \hline
$(R_{y})^{2}$&(-1)$^{p_2}$&$\eta_\bss\eta_{\bss\cs}$&$\omega_{R_y}$\\ \hline
$(\cs)^{6}=I^2$&(-1)$^{p_1+p_3}$&$\eta_\cs$&$\omega_\cs$\\ \hline
$T_1^{-1}\bst^{-1}T_1\bst$&1&1&1\\ \hline
$T_2^{-1}\bst^{-1}T_2\bst$&1&1&1\\ \hline
$R_y^{-1}\bst^{-1}R_y\bst$&(-1)$^{p_2}$&$\eta_{\bss\bst}\eta_{\cs\bst}$&$\omega_{R_y\bst}$\\ \hline
$R_x^{-1}\bst^{-1}R_x\bst$&(-1)$^{p_2+p_3}$&$\eta_{\bss\bst}$&$\omega_{R_x\bst}$\\ \hline
$\bst^2$&-1&-1&-1\\ \hline
\end{tabular}
\caption{Kagome lattice: symmetry fractionalization class $(\mbz_2)^6\subset\mathcal{H}^2(P6mm\times Z_2^\bst,\mbz_2)$ of spinons in half-integer-spin $Z_2$ spin liquids, and their realizations in $S=1/2$ Schwinger-boson (SB) and Abrikosov-fermion (AF) representations, following the convention of \Ref{Wang2006,Lu2011}. The fractionalization classes from 2nd group cohomology\cite{Qi2015} are labeled by six $Z_2$-valued integers $\omega=\pm1$. We show that a $Z_2$ spin liquid is gapless if violating any of the following conditions: $\omega_{R_x}=\omega_{R_x\bst}$, $\omega_{R_y}=\omega_{R_y\bst}$ and $\omega_{T_1T_2}\omega_{\cs}=\omega_{R_x}\omega_{R_y}$. This leads to only $2^3=8$ distinct gapped $Z_2$ spin liquids, all realized in both SB\cite{Sachdev1992,Wang2006} and AF\cite{Lu2011} representations. There are 12 symmetry protected gapless \ztsl s in the AF construction.}
\label{tab:kagome}
\end{table}

\begin{figure}
\includegraphics[width=0.9\columnwidth]{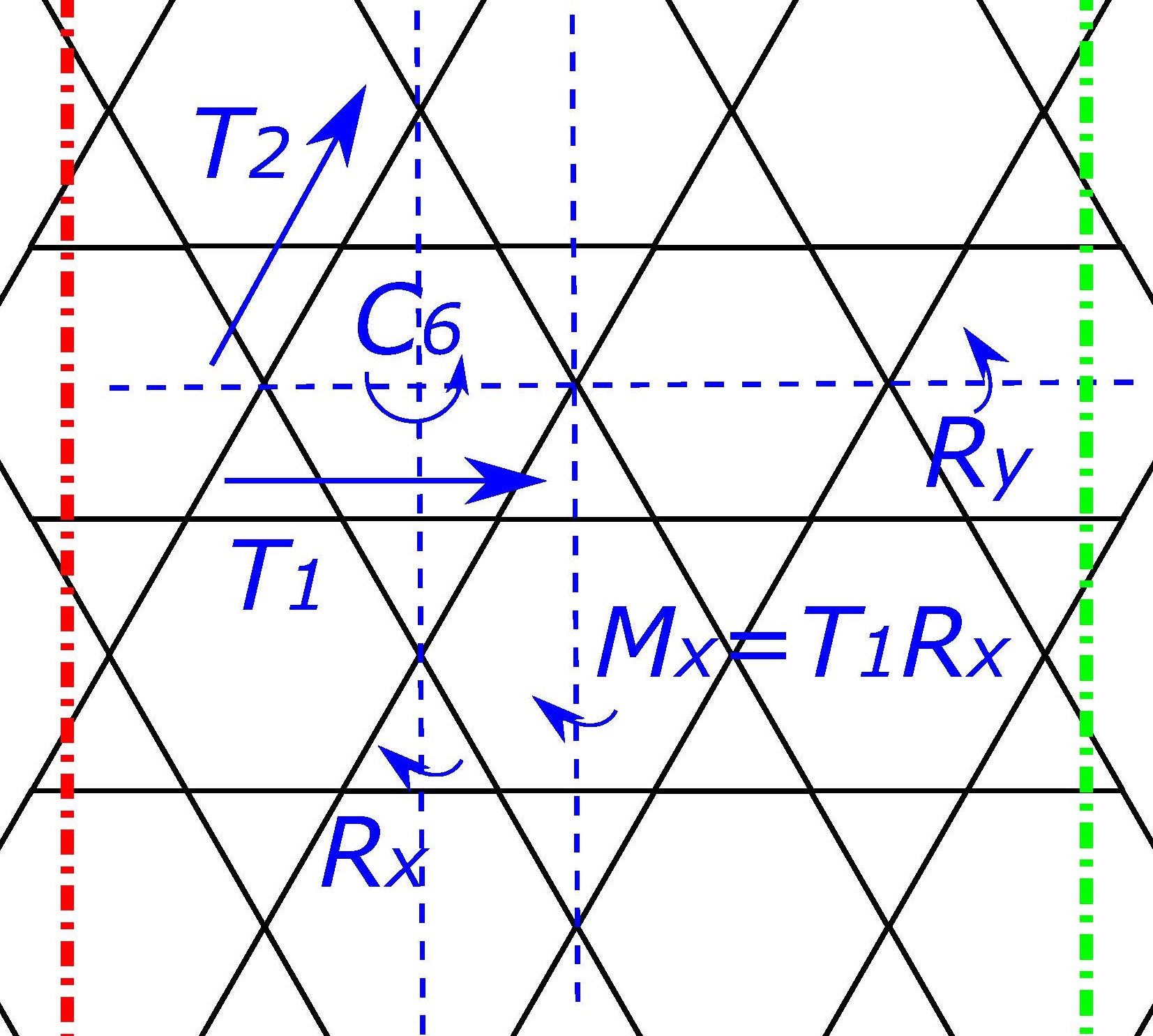}
\caption{(color online) Space group ($P6mm$) symmetries of kagome lattice, generated by two translations $T_{1,2}$, mirror reflection $R_y$ and hexagon-centered 6-fold rotation $\cs$. Two extra mirror symmetries are introduced as $R_x=(\cs)^3\bss$ and $M_x=T_1R_x$. Two mirror planes, $R_y$ and $M_x$, intersect at a lattice site. Red and green dotted dash lines denote two entanglement cuts of the lattice, related by mirror $M_x$ or inversion $M_xR_y$. }
\label{fig:kagome}
\end{figure}

\section{Gapped and gapless $Z_2$ spin liquids on the triangular lattice}\label{TRIANGLE}

As shown in FIG. \ref{fig:triangle}, the space group $P6mm$ of triangular lattice is generated by translations $T_{1,2}$, mirror reflection $\bss$ and 6-fold site-centered rotation $\cs$. The symmetry fractionalization class of spinons is classified\cite{Zheng2015} as $(\mbz_2)^6\subset\mathcal{H}^2(P6mm\times Z_2^\bst,\mbz_2)$, labeled by 6 $Z_2$-valued coefficients $\omega=\pm1$ in the last column of TABLE \ref{tab:triangle}. Again only a part of these \ztsl s are realized in spin-$\frac12$ ($2S=1$) parton constructions, leading to $2^3=8$ Schwinger-boson (SB) states\cite{Wang2006} and 20 Abrikosov-fermion (AF) states\cite{Lu2016,Zheng2015}. In fermionic parton construction (\ref{general parton construction}) of spin-$3/2$ ($2S=3$) systems, \Ref{Zheng2015} showed that all $2^6=64$ symmetry fractionalization classes of fermionic spinons can be realized.

Now we utilize the gapless criteria (\ref{criterion 1: two mirrors}), (\ref{criterion 2:mirror+TRS}) and (\ref{criterion 3:n=odd}) to identify gapless \ztsl s among all states. First, applying criterion (\ref{criterion 1: two mirrors}) to two perpendicular mirror planes $\bss$ and $R\equiv(\cs)^3\bss$ in FIG. \ref{fig:triangle} leads to one gapless condition:
\bea\label{triag:gapless:two mirror}
\frac{\omega(\bss,R)}{\omega({R,\bss})}=\frac{\omega(C_2,C_2)}{\omega(\bss,\bss)\cdot\omega(R,R)}=\omega_{\bss}\omega_R\omega_{\cs}=-1.
\eea
where we defined 2-fold rotation (or inversion) around a site
\bea
C_2\equiv\big(\cs\big)^3=\bss\cdot R\notag
\eea
Next, applying criterion (\ref{criterion 2:mirror+TRS}) to mirror plane $\bss$ and $R$ leads to two gapless conditions on the fractionalization class in TABLE \ref{tab:triangle}:
\bea\label{triag:gapless:sigma}
&\omega_{\bss}\omega_{\bss\bst}=-1\\
&\omega_{R}\omega_{R\bst}=-1\label{triag:gapless:R}
\eea
Finally, criterion (\ref{criterion 3:n=odd}) for site-centered $\cs$ rotation also leads to another gapless condition
\bea
\notag\frac{\omega(C_6,\bst)}{\omega(\bst,C_6)}=\omega_{\bss\bst}\cdot\omega_{R\bst}=-\omega_\cs.
\eea
This condition, however, is not independent of the above gapless conditions (\ref{triag:gapless:two mirror})-(\ref{triag:gapless:R}). Any spinon symmetry fractionalization class $\{\omega(g,h)\}$ that satisfies at least one of conditions (\ref{triag:gapless:two mirror})-(\ref{triag:gapless:R}) will inevitably lead to a gapless spectrum in the \ztsl.

This means any gapped symmetric \ztsl~on the triangular lattice must satisfy following ``gap condition''
\bea\label{triag:gap}
\omega_{\bss}\omega_R\omega_{\cs}=\omega_{\bss}\omega_{\bss\bst}=\omega_{R}\omega_{R\bst}=1.
\eea
on the fractionalizaton class of either fermionic or bosonic spinons.

For all 8 Schwinger-boson states\cite{Wang2006} listed in 2nd column of TABLE \ref{tab:triangle}, it's straightforward to see that ``gap conditions'' (\ref{triag:gap}) are automatically fulfilled, since the fractionalization class $\{\omega^b(g,h)\}$ of bosonic spinons $\{b_{{\bf r},\alpha}\}$ is given by
\bea
\notag&\omega^b_\bss=\omega^b_{\bss\bst}=(-1)^{p_2},~~\omega^b_R=\omega^b_{R\bst}=(-1)^{p_2+p_3},\\
&\omega^b_\cs=(-1)^{p_3}.
\eea
in the SB representation.

On the other hand, for the 20 AF states\cite{Lu2016} listed in the 3rd column of TABLE \ref{tab:triangle}, the gap condition (\ref{triag:gap}) leads to
\bea\label{triag:gap:AF}
\eta_\bss=\eta_{\bss\bst}=\eta_{\bss\cs}\eta_\cs,~~\eta_{\cs}=\eta_{\cs\bst}.
\eea
The symmetry fractionalization class $\{\omega^f(g,h)\}$ of fermionic spinons $\{f_{{\bf r},\alpha}\}$ is given by
\bea
\notag&\omega^f_\bss=\eta_\bss,~~\omega^f_{\bss\bst}=\eta_{\bss\bst},~~\omega_\cs=\eta_\cs,\\
&\omega^f_R=\eta_{\bss\cs},~~\omega^f_{R\bst}=\eta_{\cs\bst}\eta_{\bss\bst}.
\eea
in the AF representation. Only 8 states among all 20 satisfy the above gap conditions (\ref{triag:gap:AF}), and lead to gapped symmetric \ztsl s on triangular lattice.

Similar to the square lattice case, these 8 AF states have a one-to-one correspondence with the 8 SB states on triangular lattice. As shown in \Ref{Lu2016}, the duality between SB and AF states are established in terms of their bosonic/fermionic fractionalization classes:
\bea
&\notag\eta_{12}=(-1)^{p_1+1},~~\eta_{\cs}=\eta_{\cs\bst}=(-1)^{p_3},\\
&\eta_\bss=\eta_{\bss\bst}=\eta_\cs\eta_{\bss\cs}=(-1)^{p_2+1}.
\eea
Therefore they both describe the 8 distinct gapped symmetric \ztsl s on the triangular lattice. \\

In addition to the 8 gapped \ztsl s, there are 12 other mean-field states in the AF representation of a spin-$\frac12$ system on triangular lattice\cite{Lu2016,Zheng2015}. They all satisfy at least one of gapless conditions (\ref{triag:gapless:two mirror})-(\ref{triag:gapless:R}). As a result, there are 12 symmetry protected gapless \ztsl s realized in the AF representation, which are robust against any symmetry-preserving perturbations on the triangular lattice.
%
%

\section{Gapped and gapless $Z_2$ spin liquids on the kagome lattice}\label{KAGOME}

Kagome lattice share the same space group $P6mm$ with triangular lattice, generated by translations $T_{1,2}$, mirror reflection $R_y$ and hexagon-centered rotation $\cs$, as shown in FIG. \ref{fig:kagome}. Therefore spinon fractionalization class on kagome lattice is also classified\cite{Qi2015} by $(\mbz_2)^6\subset\mathcal{H}^2(P6mm\times Z_2^\bst,\mbz_2)$, characterized by 6 $Z_2$-valued coefficients $\omega=\pm1$ in the last column of TABLE \ref{tab:kagome}. In the parton constructions of a spin-$\frac12$ ($2S=1$) system on the kagome lattice, Schwinger-boson (SB) representation leads to $2^3=8$ different \ztsl~states\cite{Wang2006}, while Abrikosov-fermion (AF) representation gives rise to 20 different states\cite{Lu2011}. Below we utilize gapless criteria (\ref{criterion 1: two mirrors}), (\ref{criterion 2:mirror+TRS}) and (\ref{criterion 3:n=odd}) to identify gapped and gapless \ztsl s among them.

According to the 1st criterion (\ref{criterion 1: two mirrors}), two perpendicular mirror planes $M_x=T_1R_x$ and $R_y$ in FIG. \ref{fig:kagome} intersect at a lattice site, leading to the following gapless condition on the spinon fractionalization class
\bea
&\notag\frac{\omega(M_x,R_y)}{\omega(R_y,M_x)}=\frac{\omega_{T_1,R_y}}{\omega_{R_y,T_1}}\cdot\frac{\omega(R_x,R_y)}{\omega(R_y,R_x)}=\\
&\omega_{T_1T_2}\omega_{R_x}\omega_{R_y}\omega_{\cs}=-1.\label{kagome:gapless:two mirror}
\eea
Next, applying 2nd criterion (\ref{criterion 2:mirror+TRS}) to mirror reflections $R_x$ and $R_y$ separately leads to two gapless conditions:
\bea\label{kagome:gapless:R_x}
&\omega_{R_x}\omega_{R_x\bst}=-1\\
&\omega_{R_y}\omega_{R_y\bst}=-1\label{kagome:gapless:R_y}
\eea
Finally, applying 3rd criterion (\ref{criterion 3:n=odd}) to site-centered 2-fold rotation or inversion
\bea
\notag I\equiv M_xR_y=T_1R_xR_y=T_1\big(\cs\big)^3
\eea
lead to another gapless condition
\bea
\notag\frac{\omega(I,\bst)}{\omega(\bst,I)}=\omega_{R_x\bst}\omega_{R_y\bst}\\
\notag=-\omega(I,I)=-\omega_\cs\omega_{T_1T_2}
\eea
Again, it's straightforward to show this condition is not independent of the above 3 gapless conditions (\ref{kagome:gapless:two mirror})-(\ref{kagome:gapless:R_y}). Satisfying one or more of these 3 gapless conditions necessarily result in a stable gapless \ztsl.

In order to achieve a gapped symmetric \ztsl~on the kagome lattice, the symmetry fractionalization class of both bosonic and fermionic spinons hence  must obey the following gap condition
\bea\label{kagome:gap}
\omega_{T_1T_2}\omega_{R_x}\omega_{R_y}\omega_{\cs}=\omega_{R_x}\omega_{R_x\bst}=\omega_{R_y}\omega_{R_y\bst}=1.
\eea
Similar to square and triangular lattices, as shown in 2nd column in TABLE \ref{tab:kagome}, all 8 spin-$\frac12$ Schwinger-boson \ztsl~states on kagome lattice\cite{Wang2006} automatically satisfy the above gap conditions (\ref{kagome:gap}). In particular, the fractionalization class of bosonic spinons $\{b_{{\bf r},\alpha}\}$ is given by
\bea
&\notag\omega^b_{T_1T_2}=(-1)^{p_1},~~\omega^b_{R_y}=\omega^b_{R_y\bst}=(-1)^{p_2},\\
&\omega^b_\cs=(-1)^{p_1+p_3},~~\omega^b_{R_x}=\omega^b_{R_x\bst}=(-1)^{p_2+p_3}.
\eea
in the SB representation. All 8 SB states are gaped symmetric \ztsl s.

On the other hand, for AF \ztsl~states in 3rd column of TABLE \ref{tab:kagome}, fractionalization class of fermionic spinons $\{f_{{\bf r},\alpha}\}$ is given by
\bea
&\notag\omega^f_{T_1T_2}=\eta_{12},~~\omega^f_{R_x}=\eta_\bss,~~\omega^f_{R_x\bst}=\eta_{\bss\bst},\\
&\omega^f_\cs=\eta_\cs,~~\omega^b_{R_y}=\eta_\bss\eta_{\bss\cs},~~\omega^f_{R_y\bst}=\eta_{\bss\bst}\eta_{\cs\bst}.
\eea
in the AF representation. The gap condition (\ref{kagome:gap}) hence provides the following constraints on the fractionalization class of fermonic spinons:
\bea
\eta_{12}\eta_{\cs}\eta_{\bss\cs}=\eta_\bss\eta_{\bss\bst}=\eta_{\bss\cs}\eta_{\cs\bst}=1.
\eea
Among all 20 AF states\cite{Lu2011}, only 8 fulfill this constraint and lead to gapped symmetric \ztsl s.

Analogous to the case of square and triangular lattices, there is a one-to-one correspondence between the 8 SB states and 8 gapped AF states, established by the unique vision fractionalization class in a symmetric \ztsl~on the square lattice. As shown in \Ref{Lu2017}, the duality between 8 SB and 8 AF states is given by
\bea
&\notag\eta_{12}=(-1)^{p_1+1},~~\eta_\bss=\eta_{\bss\bst}=(-1)^{p_2+p_3+1},\\
&\eta_{\cs\bst}=\eta_{\bss\cs}=(-1)^{p_2},~~\eta_\cs=(-1)^{p_1+p_3+1}.
\eea
Hence there are 8 gapped symmetric \ztsl s on the kagome lattice, realizable in both SB and AF constructions.\\

In addition to the 8 gapped states, the other 12 states in AF representation all satisfy at least one of gapless conditions (\ref{kagome:gapless:two mirror})-(\ref{kagome:gapless:R_y}). They realize 12 different gapless \ztsl s on the kagome lattice, stable against any symmetry-preserving perturbations.

%

\section{Discussions}\label{SUMMARY}

In this work we establish a class of gapless topological phases in half-integer spin systems, which hosts symmetry-protected low-energy bulk excitations stable against any symmetric perturbations. These phases require time reversal, certain space group and at least $U(1)$ spin rotational symmetries. The low-energy dynamics of these phases are described by a (gapped) dynamical $Z_2$ gauge field coupled to gapless mater fields, half-integer ``spinons'' that transform as a Kramers doublet under time reversal symmetry. Therefore these phases are dubbed ``gapless \ztsl s''.

In the framework of parton constructions for half-integer spin systems, these gapless \ztsl s are constructed as spin-conserving pair superfluids of spinons (or partons). We establish three criteria (\ref{criterion 1: two mirrors}), (\ref{criterion 2:mirror+TRS}), (\ref{criterion 3:n=odd})-(\ref{criterion 3:n=even}) regarding the spatial (and time reversal) symmetry implementations on spinons. These criteria forbid a gapped spinon pair superfluid for arbitrary interactions between spinons. The non-perturbative proof for gapless pair superfluids of interacting spinons is achieved by using properties of Schmidt decomposition of a generic SRE state\cite{Pollmann2010,Chen2011a,Watanabe2015}. These criteria naturally lead to non-perturbative necessary conditions for a gapped pair superfluid of interacting (hard-core) spinons.

We apply these gapless criteria to symmetric \ztsl s on square, triangular and kagome lattices, and found that the only gapped \ztsl s that can be realized on these lattices are the Schwinger-boson states\cite{Sachdev1992,Wang2006,Yang2016}. There are only $2^6=64$ gapped \ztsl s on square lattice, and $2^3=8$ gapped states on triangular or kagome lattice. All these states can also be realized in the Abrikosov-fermion representation\cite{Wen2002,Lu2016,Lu2011}. In addition to these gapped states, there is also a large number of symmetry protected gapless spin-$1/2$ \ztsl s in Abrikosov-fermion construction: 96 on square lattice, 12 on kagome or triangular lattice. They are stable against any symmetry-preserving perturbations.

One thing worth mentioning is that these non-perturbative gapless criteria for a pair superfluid of interacting hard-core bosons/fermions hold quite generally, regardless of the topology of the ground state. In other words, they can be applied to a trivial pair superfluid or a topological pair superfluid with protected edge states\cite{Schnyder2008,Qi2011}. Once coupled to a dynamical $Z_2$ gauge field, a topological pair superfluid can describe other topological orders (such as double semion theory\cite{Levin2005}) other than \ztsl s. One future direction is to apply these criteria to other interacting topological phases.\\

\emph{Note added:} Upon completion of this work we became aware of Qi and Cheng's work\cite{Qi2016}, which classified all gapped symmetric \ztsl s on kagome/triangular lattice in the generic formalism of symmetry fractionalization. Their results agree with ours from parton construction on kagome/triangular lattices.

\acknowledgements{I thank Ashvin Vishwanath and Mike Zaletel for previous collaborations, as well as Yang Qi and Meng Cheng for sending me their draft. This work is supported by startup funds at Ohio State University.}



%

\end{document}